\documentclass[pdflatex,sn-basic,Numbered]{sn-jnl}

\usepackage{graphicx}
\usepackage{multirow}
\usepackage{amsmath,amssymb,amsfonts}
\usepackage{booktabs}
\usepackage{subcaption}
\usepackage{placeins}
\usepackage{xcolor}
\usepackage{textcomp}

\graphicspath{{figures/}}

\begin{document}

\title[Personalised Urban Accessibility Assessment]{An Interactive Framework for Personalised Urban Accessibility Assessment}

\author*[1]{\fnm{Prabhanjana} \sur{Ghuriki}}
\email{g.prabhanjana@bsccsh.christuniversity.in}

\author[1]{\fnm{S} \sur{Chanti}}
\email{schanti14@gmail.com}

\affil*[1]{\orgdiv{Department of Computer Science},
\orgname{Christ University},
\orgaddress{\city{Bengaluru}, \country{India}}}

\abstract{Publicly available walkability tools are easy to use but place-based:
they assign the same value to everyone sharing a location, computed from fixed
category weights and category definitions calibrated on North American service
landscapes. Accessibility research has long held that accessibility varies
among individuals at the same location, and has accordingly developed preference-
and person-sensitive measures; those measures are run as analyst-operated
geographic information system workflows, out of reach of the residents whose
preferences they represent. We present a deployed framework in which the
preference structure is supplied by the user at query time. A user
states which amenity categories matter, how much each matters, which are
non-negotiable, which to avoid, which substitute for one another, and how far
they will walk for each, and the map updates accordingly. This is possible
because a rectangular travel impedance reduces the measure to the number of
amenities reachable within a user-set threshold, which can be precomputed via
shortest-path routing, leaving only arithmetic at query time. The score also
stays interpretable: it is bounded, and each category's contribution can be read
on its own. We evaluate the framework across Bengaluru, reporting its
sensitivity to discretisation choices and comparing it against existing
measures using identical inputs. Configuration moves ward rankings considerably
more than the choice of aggregation does.}

\keywords{urban accessibility; personalised accessibility; cumulative opportunity accessibility; 15-minute city; multi-criteria decision analysis}

\maketitle

\section{Introduction}
\label{sec:intro}

\subsection{The urban accessibility challenge in Indian cities}
\label{sec:intro_context}

Urban accessibility is the ease with which residents can reach services and
opportunities. \citet{jacobs1961death} established the conceptual foundation for walkable urban
environments, which subsequent decades of research have attempted to quantify as
an attribute of location.

European cities such as Paris and Barcelona have implemented proximity-based
planning principles \citep{NIEUWENHUIJSENbarcelona,nattererparis}. Indian cities
face conditions that differ substantially: extreme population densities, acute
socioeconomic heterogeneity, rapid and often informal urban growth, and the
co-existence of different service provision models within the same
neighbourhood \citep{shaw2012urban}. Populations with divergent service needs
and economic constraints are routinely co-located, so a single averaged
assessment is least informative where populations are most mixed. Many
major Indian cities also contend with uneven service distribution and
inequitable access to opportunities across income groups
\citep{sudhira2007urban}, and walking is common but poorly provided for:
\citet{rahul2021acceptable} report that the proportion of roads with pedestrian
footpaths in most Indian cities is around thirty per cent, and that pedestrians
accounted for 44\,\% of total road accidents in Bengaluru.

The tools a resident can actually consult reflect none of this. Widely available
public walkability indices are, for the most part, either unavailable outside
North America and Europe or built on assumptions calibrated there. Where such a
method could in principle be applied in an Indian city, its category set often
cannot: an index whose grocery category is defined by supermarkets measures
something different in a city where daily provisioning runs through
neighbourhood kirana stores, and one whose weights encode North American
trip-making will misrank neighbourhoods whose residents travel differently.
\citet{steiniger2019walkmyplace} report exactly this in Chile, where Walk Score
omitted locally important destinations such as street markets and weighted
eating and drinking out more heavily than Chilean travel patterns warrant.
These tools are also place-based, assigning the same value to everyone sharing a
location.

\subsection{The fifteen-minute city as a conceptual foundation}
\label{sec:intro_15min}

The fifteen-minute city concept, proposed by Carlos Moreno and later articulated
in \citet{smartcities}, prioritises proximity over mobility, enabling residents
to meet daily needs through active transport rather than private vehicle
dependency, and has been formalised as accessibility by proximity
\citep{BUTTNER2024100095}. Its core proposition, that a walkable neighbourhood is
one where daily needs fall inside a short walk, is the one this framework
operationalises.

\subsection{Contribution}
\label{sec:intro_contribution}

Producing an accessibility surface from a preference-sensitive measure is an
analyst's task (Section~\ref{sec:lit_personbased}): it requires data, software,
routing and time, and changing a single weight means running it again. Public tools
invert the trade, offering immediacy and almost no preference expression. What
this work supplies is the missing delivery: preference specification an ordinary
user can perform at urban scale and at interactive speed. Its
contributions are:

\begin{enumerate}
\item \emph{A preference model.} A user-configurable accessibility operator
supporting category weights, user-declared substitution groups, negative
preferences, per-category non-compensatory constraints, per-category travel-time
thresholds and adjustable diminishing returns, bounded and decomposable by
category so that a score can be read as well as computed.
\item \emph{A computational factorisation.} A rectangular travel impedance
collapses the opportunity summation to threshold-specific counts that can be
precomputed once by shortest-path routing. This substitution is what converts a
batch measure into an interactive one; every user control then acts on stored
integers by arithmetic alone.
\item \emph{An empirical demonstration.} Across the 198 wards of Bengaluru, user
configuration moves ward rankings substantially further than the choice of
aggregation does, while the threshold approximation stays close to the
established formulation it extends. We report sensitivity to every
discretisation parameter, comparison against non-personalised baselines and
against aggregations given identical preference data, and the coverage limits of
open amenity data in this setting.
\end{enumerate}

\section{Literature Review}
\label{sec:lit}

\subsection{Operationalising the fifteen-minute framing}
\label{sec:lit_15min}

Translating the fifteen-minute city from a planning principle into a measurement
requires fixing a threshold, and the literature is less settled on that number
than the name suggests. \citet{BUTTNER2024100095} treat proximity as the
organising concept rather than fifteen minutes as a constant, and surveys of
European implementation practice record considerable variation in the catchment
actually adopted \citep{buttner2024mapping}.

Evidence from Indian cities suggests the appropriate threshold is shorter and
more variable than the framing implies. \citet{rahul2021acceptable}, drawing on
a household travel survey of 2\,400 households in Bengaluru, report acceptable
walking distances of 996\,m for work trips but 666\,m for pick-up and drop-off,
360\,m for recreational, medical and social trips, and 263\,m for personal and
household business. At ordinary walking speeds the shortest of these corresponds
to roughly three minutes rather than fifteen. Trip purpose is also the dominant
axis of variation in that survey: workers' work and non-work trips differ at
$Z = 10.96$.

Two implications follow. The threshold is a property of the trip and the
traveller as much as of the city, so a single constant applied to all categories
and all users discards available information. And a concept calibrated where
pedestrian provision is continuous does not transfer unexamined to a setting
where roughly thirty per cent of roads carry footpaths. Both observations argue
for treating the threshold as a quantity the user sets rather than one the tool
assumes.

\subsection{Place-based and person-based accessibility}
\label{sec:lit_personbased}

\citet{hansen1959} introduced gravity-based models
incorporating both distance decay and destination attractiveness, providing
mathematical structure for balancing distance deterrence against destination
appeal, at the cost of explicit assumptions about decay parameters and
attractiveness weights \citep{geurs2004}.
\citet{ingram1971} distinguished relative accessibility, between a pair of
points, from integral accessibility, between one point and all others; a
distinction that still underlies most operational forms.

The distinction that matters here is between place-based and person-based
measures, which \citet{neutens2010} set out directly: place-based measures
evaluate the spatial separation between services and a single reference
location, while person-based measures rest on an individual's activity schedule
and space--time constraints. Comparing four place-based and six person-based
measures of urban service delivery, they find substantial differences between
the two families and within the person-based family itself.
\citet{geurs2004} locate the same division in a four-part taxonomy of
infrastructure-, location-, person- and utility-based measures, in which the
person-based component individualises by constraint, through the commitments an
individual holds and the time available between them, following the space--time
prism tradition operationalised by \citet{kwan1998space}, and the utility-based
component individualises by valuation.

Individual preference and perception enter most directly through the perceived
accessibility literature. \citet{pot2021perceived} distinguish perceived
accessibility from accessibility calculated on spatial data and set out why the
two diverge. \citet{ijgi13120450} classify measures into those built on
physical separation between locations and those capturing individuals'
preferences and attitudes toward reaching goods, services and activities, and
argue that planning needs both. \citet{CURL20113}, interviewing local authority
officers responsible for accessibility planning in England, find that modelled
accessibility does not readily translate into the improvements those
practitioners regard as reducing transport disadvantage. That accessibility
differs between people who share a location is therefore long settled, as are
the axes along which it differs: schedule and constraint on one hand, valuation
and perception on the other.

The framework presented here individualises along the second axis while
retaining place-based form. Its output is a score attached to a location,
computed from what that location can reach; the preference structure producing
that score, weights, substitution groups, constraints and thresholds, is
supplied by the user at query time rather than by an analyst at model-build
time. The computational design of Section~\ref{sec:method_arch} exists to make
that possible at interactive speed.

\subsection{Cumulative opportunity measures and diminishing returns}
\label{sec:lit_cumulative}

Modern accessibility assessment has converged around cumulative opportunity
measures, which quantify accessibility as the number of opportunities reachable
within a specified travel time. The appeal of this container-based approach is
interpretability: accessibility becomes a count of services reachable within a
fixed travel-time or distance budget, the budget itself being a choice the
analyst makes. \citet{xu2020deconstructing} show how strongly such counts depend on the
joint distribution of facilities and population, and \citet{neutens2010} caution
that measures of this family can diverge sharply from one another, and that
measures expressing the number of reachable opportunities differ systematically
from those expressing how desirable that set is.

The Walk Score index applies this logic, evaluating neighbourhoods by counting
amenities in thirteen categories within walking distance and aggregating them
into a single score \citep{carr2010walk}. Both the thirteen categories and the
rule assigning a business to one of them are set by the vendor and unpublished,
as is the weight set, which is why the reconstruction of
Section~\ref{sec:results_walkscore} must supply its own category mapping.
\citet{Brownetal}, reviewing its use
across the health and planning literatures, note that it is frequently treated
as a validated measure of walkability despite being a proprietary proximity
score whose transferability outside North America is largely untested.

Recent fifteen-minute city research retains cumulative logic while extending the
spatial analysis. The 15-Minute City Score Toolkit scores a hexagonal cell by how
many of eight essential service categories are reachable on foot within fifteen
minutes, weighting each by an exponential distance decay function and combining
them into a single composite \citep{albashir_2024_14231533}. Its elaboration is in
spatial detail rather than in the aggregation: an isochrone for every minute up
to fifteen, at a user-selected hexagon resolution.
\citet{JIANG2024103197} quantify liveable areas in Auckland's suburban
neighbourhoods using hexagonal grid analysis over six service categories, and
their one-factor-at-a-time experiment clarifies how individual services
contribute by varying one input at a time on common data. Where supply capacity
and competing demand matter, two-step floating catchment area methods
\citep{luo2003measures} compute a provider-to-population ratio within each
catchment and sum those ratios at each origin; they are the standard tool for
facilities with finite capacity, such as clinics or school places.

\citet{roper2023incorporating} advance the field through their WalkTHERE index,
an open-source and reproducible approach based on multi-activity accessibility
rather than single-purpose trips. Categories are drawn from household travel
survey trip purposes and weighted by trip frequency. The index employs
continuous distance decay through exponential impedance functions in place of
hard thresholds, and incorporates diminishing marginal utility of successive
opportunities, so that the first few accessible shops, jobs or schools matter
more than the twentieth or thirtieth. The principle is stated formally by
\citet{Levinson_Wu_2020}, who discount an opportunity count by a negative
exponential in the number reachable for exactly this reason, and Walk Score, as
formalised by \citet{dai2023walkability}, discounts successive opportunities
too.

\subsection{Preference, substitution and multi-criteria aggregation}
\label{sec:lit_mcda}

Public indices apply uniform category weights to all residents, assuming
identical service needs irrespective of demographic characteristics or cultural
context \citep{CURL20113}. Walk Score illustrates this through fixed weights
applied to amenity categories \citep{carr2010walk}; its full weight set is in
any case patented and unpublished, so even its assumptions are not open to
inspection. In cities where households with vastly different incomes occupy the
same neighbourhood, uniform treatment is particularly consequential.

The same rigidity applies to the shape of the measure and not only its weights.
No public index offers control over the rate of discounting, or over which
categories it applies to. \citet{Levinson_Wu_2020} identify the related
assumption: summing weighted opportunity types presumes those types are
substitutes, when some are complements. Their example is that restaurants can
substitute for grocery stores while jobs and shops complement one another, and
their notation carries a matrix of activity substitutability and
complementarity for the purpose. Which pairs substitute is also, we would add, a
personal matter. Fast food outlets and restaurants may be substitutes for some
residents but not for those managing tight budgets; banks and ATMs may
substitute for basic transactions but not for someone requiring full-service
branches. Simple weighted summation is therefore misleading when applied across
diverse populations. Frameworks that bundle destination types rather than scoring each independently
exist \citep{klumpenhouwer2021bundling}, but the bundles are composed by the
analyst.

Combining several criteria into a single score for a set of spatial alternatives
is the subject of GIS-based multi-criteria decision analysis
\citep{malczewski2006gis}. In that vocabulary, spatial units are alternatives,
amenity categories are criteria, and the composite is an aggregation. Weighted
linear combination is the most common aggregation; ordered weighted averaging
\citep{yager1988owa} allows movement between fully compensatory aggregation,
where abundance in one criterion offsets absence in another, and conjunctive
aggregation, where every criterion must be satisfied. This vocabulary supplies
the per-criterion hard constraint used later to express a non-negotiable need.

Multi-criteria approaches are, however, run by analysts: producing one surface
means assembling network data, routing, intersecting and aggregating, and
changing a single weight means running it again. Elicitation compounds the
problem, since techniques such as the analytic hierarchy process require
pairwise comparisons that scale quadratically with the number of criteria, which
for fifty amenity categories is impractical for a non-specialist.

\subsection{Publicly available tools}
\label{sec:lit_tools}

Table~\ref{tab:tools} summarises the tools a non-specialist can use directly,
and the pattern across them is consistent. Walk Score \citep{carr2010walk} is
the most widely used and asks nothing of the user, but its category set and
weights are fixed. The EPA National Walkability Index
\citep{epa2021walkability} is a static dataset of ranked block groups. The
15-Minute City Score Toolkit \citep{albashir_2024_14231533} is configurable, but
configuration means editing and running code, which places it with the research
methods rather than the public tools.

\begin{table}[htbp]
\caption{Publicly available accessibility tools. ``Weighting'' indicates whether
a user can express that one category matters more than another.}
\label{tab:tools}
\footnotesize
\begin{tabular*}{\textwidth}{@{\extracolsep{\fill}}llll@{}}
\toprule
\textbf{Tool} & \textbf{Output} & \textbf{Categories} & \textbf{Weighting} \\
\midrule
Walk Score \citep{carr2010walk} & score 0--100 & fixed & fixed \\
EPA Walkability Index \citep{epa2021walkability} & decile rank & fixed & fixed \\
Close \citep{henry2024close} & travel-time bands & user-set & none \\
15min City Score Toolkit \citep{albashir_2024_14231533} & score & via code & via code \\
\bottomrule
\end{tabular*}
\end{table}

Close \citep{henry2024close} comes nearest to bridging the divide. Users select
the destinations that matter to them and choose a travel mode for each, and the
map colours every block by travel time. Its output is a travel-time surface
rather than a composite score, so blocks are shaded in time bands rather than
rated. And it colours each block by the travel time to the \emph{furthest}
selected destination, treating all chosen destinations as equally necessary,
with no facility for expressing that one matters more than another; in the
vocabulary of the preceding section, a conjunctive operator without weighting.

\subsection{The implementation gap}
\label{sec:lit_gap}

Each literature supplies half of what is needed. The research measures reviewed
above admit per-criterion hard constraints, explicit treatments of
substitutability and purpose-specific thresholds, but are operated as offline
analyses; the tools of Table~\ref{tab:tools} are immediate and expose almost
none of that structure. Consider a household deciding where to live that walks its
children to school, wants a pharmacy within a few minutes, treats any of several
grocers as interchangeable, and would rather not be beside a row of bars. Every
one of those statements is expressible in the vocabulary above, and none can be
entered into a tool the household can operate.

\section{Methodology}
\label{sec:method}

\subsection{Conceptual framework}
\label{sec:method_concept}

The framework takes an established treatment of diminishing returns
\citep{roper2023incorporating} and adds the personalisation controls and the
two-stage architecture that let a user-configured score be evaluated in real
time.

The framework applies decay to the \emph{count} of overlapping catchments within
a hard, user-selected travel-time threshold. This pairs a sharp and intuitive
proximity threshold, under which an amenity is either reachable within the
chosen walk or it is not, with diminishing returns on \emph{redundancy}, so that
each additional reachable amenity of the same type contributes progressively
less. Keeping the two independently interpretable is what makes
the controls usable by a non-specialist and, as Section~\ref{sec:method_arch}
shows, also makes the score computable at interactive speed.

Personalisation operates through four mechanisms, each corresponding to a term
in the score. Users select which amenity categories they value and which they
would rather avoid, and for each selected category choose how strongly that
preference should count: adjusting its weight, or marking it as absolute. On the
positive side, a category may be \emph{Wanted}, contributing to the score by
degree, or \emph{Required}, acting as a filter whose absence renders a location
unsuitable regardless of what else is present. On the negative side, a category
may be marked \emph{Avoid}, deducting a saturating penalty as unwanted instances
accumulate, or \emph{Never}, its hard mirror, setting the score to zero if a
single instance is reachable. Users may group services
that are functionally interchangeable for them, so that substitutable options
saturate together rather than accumulating independently. Users set the decay
rate, controlling how quickly redundancy stops adding value. And users set the
walking-time threshold separately for each category.

\subsubsection{Mathematical formulation}
\label{sec:method_formulation}

Table~\ref{tab:notation} gives the notation. The score is defined in two parts:
a compensatory core in which every control acts by degree, and a
non-compensatory filter that overrides it.

\begin{table}[htbp]
\caption{Notation used throughout the accessibility methodology.}
\label{tab:notation}
\footnotesize
\begin{tabular*}{\textwidth}{@{\extracolsep{\fill}}cl@{}}
\toprule
\textbf{Symbol} & \textbf{Definition} \\
\midrule
$c$ & Amenity category or user-defined group index \\
$P$ & Set of groups the user wants nearby \\
$N$ & Set of groups the user would rather avoid \\
$R \subseteq P$ & Groups designated \emph{required} \\
$Z \subseteq N$ & Groups designated \emph{never} \\
$w_c$ & User-set weight for group $c$ \\
$\lambda_c$ & Decay parameter for group $c$; user-set for $c \in P$, fixed for $c \in N$ \\
$\tau_c$ & User-set travel-time threshold for group $c$, in minutes \\
$k_{g,c}(\tau_c)$ & Number of catchments of group $c$ covering cell $g$ at $\tau_c$ \\
$A_c$ & Set of all amenities in group $c$ \\
$g$ & Grid cell index \\
$\tilde{s}_g$ & Compensatory core of the cell score \\
$s_g$ & Accessibility score for grid cell $g$ \\
$G_W$ & Set of grid cells assigned to ward $W$ \\
$S_W$ & Ward-level accessibility score \\
\bottomrule
\end{tabular*}
\end{table}

\bmhead{Grid-cell score}
For grid cell $g$, category $c$ and threshold $\tau_c$, the overlapping
catchment count is
\begin{equation}
k_{g,c}(\tau_c) = \sum_{a \in A_c}
  \mathbf{1}\!\left[\frac{\mathrm{area}(C_a(\tau_c) \cap g)}{\mathrm{area}(g)}
  \geq 0.5\right],
\label{eq:grid_coverage}
\end{equation}
where $C_a(\tau_c)$ is the catchment of amenity $a$ at threshold $\tau_c$ and
$\mathbf{1}[\cdot]$ is the indicator function. This counts the catchments
covering at least half of the cell.

Selected categories fall into two sets: $P$, those the user wants nearby, and
$N$, those they would rather avoid. The compensatory core combines the user's
weights, decay parameters and thresholds as
\begin{equation}
\tilde{s}_g =
\frac{\displaystyle\sum_{c \in P} w_c \left(1 - e^{-\lambda_c k_{g,c}(\tau_c)}\right)
    - \sum_{c \in N} w_c \left(1 - e^{-\lambda_c k_{g,c}(\tau_c)}\right)}
     {\displaystyle\sum_{c \in P} w_c},
\label{eq:grid_score}
\end{equation}
and the reported score applies the hard constraints and a clamp at zero:
\begin{equation}
s_g =
\begin{cases}
0, & \text{if } \exists\, c \in R:\, k_{g,c}(\tau_c) = 0
     \ \text{ or } \ \exists\, c \in Z:\, k_{g,c}(\tau_c) > 0,\\[2pt]
\max\!\left(0,\ \tilde{s}_g\right), & \text{otherwise.}
\end{cases}
\label{eq:final_score}
\end{equation}

\bmhead{Boundedness}
Since $1 - e^{-\lambda_c k} \in [0,1)$ for $\lambda_c > 0$ and $k \geq 0$, each
wanted term is bounded above by its own weight $w_c$, so the positive sum is
strictly less than $\sum_{c \in P} w_c$ and the numerator's first term is
strictly less than 1 once divided. The avoidance sum is non-negative, so it can
only reduce the result. With the clamp, $s_g \in [0,1)$ under every
configuration. The exponential form is what supplies this: it also encodes
diminishing marginal utility as overlapping amenities accumulate.

\bmhead{Mapping controls to terms}
Every user-facing control appears in
Equations~\ref{eq:grid_score}--\ref{eq:final_score}, and each maps to one place.
Category selection determines which terms enter the sums. Grouping determines
what each $k$ counts, since a group's count is the sum of its members' counts
taken before the decay is applied, $k_{\text{group}} = \sum_{i} k_{c_i}$. The
treatment control places the group in $P$ or in $N$, and may additionally
designate it \emph{required} ($R \subseteq P$) or \emph{never} ($Z \subseteq
N$). Weighting
sets $w_c$. The ``how many nearby'' setting sets $\lambda_c$ for $c \in P$; for
$c \in N$ it is not offered and $\lambda_c$ takes the middle value, since the
question of how many one wants does not arise for something one wants none of.
The threshold sets $\tau_c$.

The two sets are deliberately asymmetric: only $P$ enters the denominator. An
avoided category therefore subtracts where it is present and does nothing where
it is absent, so naming one can lower a location's score but never raise it.
Were it to enter the denominator instead, adding an avoided category would
inflate scores everywhere the unwanted amenity happens not to be. The
denominator is zero when $P$ is empty, and $s_g$ is defined as zero in
that case: a configuration naming only things to avoid states no preference the
measure can score.

\bmhead{Hard constraints change the feasible set}
The two designations in Equation~\ref{eq:final_score} are the framework's only
non-compensatory operations, and they are mirror images of one another. A weight changes the ranking of locations within
the feasible set; \emph{required} and \emph{never} change which locations are
feasible at all. No value of $w_c$ in a weighted linear combination reproduces
either, because raising a weight cannot drive a location's score to zero on the
strength of one absent category while every other category is well served.
Section~\ref{sec:results_mcda} quantifies the consequence.

\bmhead{Relation to multi-activity accessibility}
Equation~\ref{eq:grid_score} extends the WalkTHERE formulation of
\citet{roper2023incorporating}, whose accessibility at origin $i$ is
$A_i = \sum_c W_c \sum_j U_{jc} f_c(d_{ij})$, with $W_c$ a category weight,
$f_c$ an impedance function of network distance, and $U_{jc}$ the marginal
utility of the $j$th opportunity, itself defined so that successive
opportunities contribute progressively less. Writing $O_\ell$ for the weight
that formulation assigns to the $\ell$th opportunity, the marginal-utility term
summed over the first $n$ opportunities telescopes:
\begin{equation}
\sum_{j=1}^{n}\left(e^{-\lambda\sum_{\ell=1}^{j-1}O_\ell}
                  - e^{-\lambda\sum_{\ell=1}^{j}O_\ell}\right)
= 1 - e^{-\lambda \sum_{\ell=1}^{n} O_\ell}.
\label{eq:telescope}
\end{equation}
Equation~\ref{eq:telescope} holds for any opportunity weights. It reduces to
$1 - e^{-\lambda n}$, the saturating term of Equation~\ref{eq:grid_score},
exactly when $O_\ell = 1$ for every opportunity, which is the value
\citet{roper2023incorporating} adopt for the education, shopping, errand and
recreation categories. Our count treats every instance of a category alike and
so makes the same assumption; where a source distinguishes opportunities by size
or capacity, the general form applies and a count is not sufficient.

The extension lies in the impedance. Where WalkTHERE uses a continuous negative
exponential $f_c(d) = e^{-\lambda_c d}$, we use a rectangular function
$f_c(d) = \mathbf{1}[d \leq v\tau_c]$ for walking speed $v$, with $\tau_c$ under
the user's control; in the terms of \citet{Levinson_Wu_2020}, the isochrone
measure is the special case in which the decay function is rectangular. Under a rectangular impedance the inner summation collapses to a count of
opportunities within a threshold, and a count can be computed once and stored.
The user can then change weights, decay parameters and thresholds, and the score
is recovered by arithmetic alone. A continuous impedance admits no such
factorisation, because each opportunity's contribution depends on parameters
unknown until the user supplies them. The substitution is therefore what
converts a batch measure into an interactive one, and
Section~\ref{sec:results_comparison} reports how far it moves rankings relative
to the continuous form.

\bmhead{Ward-level score}
Wards are the administrative unit at which provision is planned in BBMP, so
results are reported at ward level alongside the grid. The ward score is the
mean of its constituent cells,
\begin{equation}
S_W = \frac{1}{|G_W|}\sum_{g \in G_W} s_g .
\label{eq:ward_mean}
\end{equation}
Cells are assigned by plurality: a cell belongs to the ward covering the largest
share of its area, and only cells overlapping no ward polygon are left
unassigned. The grid covers the bounding box of BBMP and so contains 20\,155
cells, of which 11\,752 fall inside a ward; the remainder lie outside the
municipal boundary and are excluded from every ward-level statistic and masked
out of every map.

\bmhead{Interpreting the score}
Scores are intended to be read comparatively rather than as absolute measures.
As a guide, a value near 0.5 indicates baseline single-source coverage, values
around 0.75 indicate comfortable provision, and values approaching 1 indicate
redundant provision that buffers against the closure, congestion or
unavailability of any single facility, which is the concern motivating
backup-coverage formulations in facility location \citep{Fadda2024}. It follows
by design that a solitary
hospital or metro station does not saturate the score even under the fastest
decay setting.

Where a configuration includes something to avoid, the same scale applies but is
read as provision net of nuisance: a location with comfortable provision and an
unwanted neighbour lands where a location with thinner provision and no such
neighbour would. The two are not distinguishable from the composite alone, which
is a general property of any weighted aggregation and the reason the interface
reports each group's contribution alongside the total.

\subsection{Personalisation controls}
\label{sec:method_personalisation}

Each mechanism exposes a small set of discrete options rather than a continuous
control, trading fine granularity for interpretability so that a user can
express a meaningful difference in priority without calibrating a parameter.

\bmhead{Category selection and grouping}
Amenity categories may be combined into groups. Bundling destinations
is not itself new \citep{klumpenhouwer2021bundling}; what differs is who
composes the bundle and when. Setting
$k_{\text{dining}} = k_{\text{restaurant}} + k_{\text{fast food}} +
k_{\text{caf\'e}}$ asserts that these opportunities belong to one functional
opportunity set for this user, so that any one of them satisfies the need. Counts are combined before decay, so interchangeable options share saturation rather than each being credited independently. 

\bmhead{How it's treated}
A single control places a category in $P$ or $N$, in either a soft or a hard
form. \emph{Wanted} places it in $P$ as an ordinary contributor, and is the
default. \emph{Required} places it in $P$ and in $R$: the score is zero wherever
the category is unreachable, which is how a user states a non-negotiable need.
\emph{Avoid} places it in $N$, deducting a saturating penalty as unwanted
instances accumulate, so one nearby is a mild cost and several a substantial
one, while absence leaves the score untouched. \emph{Never} places it in $Z$: a
single reachable instance sets the score to zero, exactly as Required does in
reverse.

\bmhead{How much it matters}
Weighting is a separate axis, applying to whichever set the category sits in.
Standard ($w_c = 1$) treats a category at baseline importance and Emphasised
($w_c = 2$) doubles it; its contribution in $P$, its penalty in $N$, so that a
strongly disliked neighbour can outweigh a merely useful one. It applies only to
the two graded treatments. Required and Never are statements about feasibility
rather than degree, so both take $w_c = 1$, and Never contributes to neither sum
in any case.

\bmhead{How many nearby}
The decay control asks a simple question: once you have one of something, how
much does a second or third add? It uses a half-life parameterisation, where the
half-life is the number of overlapping amenities at which the benefit reaches
half its asymptotic maximum, and is presented in those terms rather than as a
rate. \emph{A few} ($\lambda = 2\ln 2 \approx 1.386$) gives a half-life of
$k = 0.5$: benefit saturates quickly and one or two nearby options largely
suffice. \emph{A handful}
($\lambda = \ln 2 \approx 0.693$) gives a half-life of $k = 1$, so each
additional amenity is worth about half the previous. \emph{A lot}
($\lambda = \ln 2/2 \approx 0.347$) gives a half-life of $k = 2$, suited to
categories where variety keeps adding utility well beyond the
first option. The control does not apply to avoided categories; their penalty
accumulates at the middle rate.

\bmhead{How far you'll walk}
Users set $\tau_c \in \{3,4,\dots,15\}$ minutes independently per category. The range extends to three minutes because, as
Section~\ref{sec:lit_15min} notes, two of the four reported acceptable distances
for Bengaluru fall below a five-minute walk.

All four controls default to a common setting: every category Wanted at Standard
weight, \emph{a handful} nearby and a nine-minute threshold, so that a user who
expresses no preference receives a plain uniform-weight accessibility surface,
and every departure from it is one the user has made deliberately. Nine minutes
sits between the acceptable distances reported for work and for non-work trips
in Section~\ref{sec:lit_15min}.

\subsection{Study area and data}
\label{sec:method_data}

Bengaluru was selected for its mix of formal and informal urban development,
which permits comparison of accessibility across sharply different neighbourhood
types, and for the relative availability of the open data this analysis
requires. The study covers all 198 wards within the administrative limits of
Bruhat Bengaluru Mahanagara Palike, India's third-largest metropolitan area.
Ward boundaries were obtained from the Urban Data Portal
\citep{urban_data_portal}, with spatial data processed in WGS84 (EPSG:4326) for
consistency with OpenStreetMap sources.

Amenity data were collected from OpenStreetMap as of July 2025
\citep{osm_france_india}, comprising 20\,790 features across 50 categories. That
choice means coverage varies by category, which Section~\ref{sec:results_data}
quantifies and Section~\ref{sec:limitations} treats as the framework's principal
constraint.

Categories are retained by an explicit rule rather than by inspection: a
category is included if it contains at least 25 amenities. Three categories
below that count were retained by exception (massage, 21; tattoo, 19; fabric,
18) because each is a distinct destination type with no substitutable sibling
among those retained. Each feature is assigned to exactly one category, the
largest candidate where a feature appears in several thematic extracts, so
features are not double counted. OpenStreetMap represents amenities as points or
polygons depending on physical size and mapping detail. Polygon amenities are
routed from their geometric centroid.

\subsection{Two-stage architecture}
\label{sec:method_arch}

Direct geometry-based evaluation of Equation~\ref{eq:grid_coverage} at query
time is too slow for interactive use. The system therefore separates
computationally intensive spatial operations, performed once offline, from the
lightweight arithmetic performed on each user interaction.

\subsubsection{Offline stage}

\bmhead{Catchment generation}
Catchments are computed by bounded shortest-path search over a pedestrian
network built from an OpenStreetMap extract, comprising 619\,696 nodes and
700\,377 edges covering 23\,367\,km of pedestrian network, with ways closed to
pedestrians excluded. Walking
speed is 80\,m\,min$^{-1}$ (4.8\,km\,h$^{-1}$), the mean sidewalk free-flow
speed reported by \citet{banerjee2018pedestrian} across international studies,
which reports a range of 65--85\,m\,min$^{-1}$. The speed fixes the distance corresponding to any threshold: at
80\,m\,min$^{-1}$ a fifteen-minute catchment reaches 1\,200\,m, consistent with
the distance generally assumed in the fifteen-minute-city literature. Routing is
performed independently at every threshold from 3 to 15 minutes.

The routed catchments were validated against isochrones produced independently
for the same origins by the Valhalla routing engine \citep{valhalla_routing},
which uses a different graph representation and derives contours from a cost
surface rather than from the set of reached nodes. Mean
intersection-over-union across sampled categories was 0.868, with disagreement
concentrated at catchment boundaries rather than in the interior.

\bmhead{Spatial discretisation}
The study area is discretised into 250\,m cells, a scale consistent with
neighbourhood-level walkability analysis \citep{BUTTNER2024100095} and coarse
enough to permit real-time interaction across the metropolitan area. Coverage is
evaluated by sampling rather than polygon intersection: each cell carries a
regular $7\times7$ lattice of points, and a cell counts as covered when at least
half its sample points are reachable within the threshold. This estimates the
same criterion as Equation~\ref{eq:grid_coverage} without requiring catchment
polygons to be constructed, and therefore without a polygon-shape parameter.

\bmhead{Sensitivity to discretisation}
Three parameters govern the discretisation, and each was varied independently
with all else held constant. Cell size is the most consequential: ward rankings
are effectively unchanged between 100\,m and 250\,m ($\rho = 0.9987$, median
rank shift 1 place of 198) and begin to degrade at 500\,m, where individual
wards move up to 39 places (Table~\ref{tab:sens_grid}). The 250\,m resolution
used throughout sits inside the stable regime at a fraction of the computational
cost of 100\,m. Ward polygons are held fixed across all three grids, so
assignment differences at ward boundaries are a property of cell size rather
than of the boundaries.

\begin{table}[htbp]
\caption{Sensitivity to grid cell size. Agreement is against the 250\,m grid at
ward level.}
\label{tab:sens_grid}
\footnotesize
\begin{tabular*}{\textwidth}{@{\extracolsep{\fill}}lrrrrr@{}}
\toprule
\textbf{Cell size} & \textbf{Cells} & \textbf{Pearson} & \textbf{Spearman} & \textbf{Median shift} & \textbf{Max shift} \\
\midrule
100\,m & 131\,350 & 0.9988 & 0.9987 & 1 & 13 \\
250\,m & 20\,155 & -- & -- & -- & -- \\
500\,m & 5\,254 & 0.9908 & 0.9898 & 3 & 39 \\
\bottomrule
\end{tabular*}
\end{table}

Varying the coverage fraction a cell must satisfy from 40\,\% to 60\,\% changes
mean scores by $\mp 5\,\%$ and leaves ward rankings almost unaffected
($r \geq 0.992$, median rank shift 1, top-20 ward overlap 20/20 in both
directions). Refining the sample lattice from $3\times3$ to $9\times9$ changes
mean scores by less than $10^{-4}$ and raises cross-resolution agreement
monotonically ($r = 0.9974$, $0.9985$, $0.9992$); residual disagreement in
individual coverage counts decays sub-linearly (12.6\,\%, 7.7\,\%, 5.4\,\%)
toward a floor set by street-network node density rather than by lattice
density, and is almost entirely $\pm 1$ in $k$. A $7\times7$ lattice is used
throughout, $9\times9$ costing 65\,\% more computation for an improvement in
agreement from 0.9985 to 0.9992.

Absolute scores therefore shift by a few per cent across all three parameters
while ward rankings barely move, which is the basis for reading scores
comparatively.

\subsubsection{Real-time stage}

For each cell and category the offline stage stores a vector of integer counts,
one per threshold. On each interaction the application retrieves the count for
every selected group at its chosen threshold, evaluates
Equations~\ref{eq:grid_score}--\ref{eq:final_score}, and where required
aggregates through Equation~\ref{eq:ward_mean}. Storage scales as
$C \times T$ integers per cell for $C$ categories and $T$ thresholds. Query cost
is one lookup and a small number of arithmetic operations per group, independent
of the number of amenities, which is what permits the map to update as the user
changes any control.

\subsection{Interactive web application}
\label{sec:method_webapp}

The methodology is implemented as a responsive web
application.\footnote{\url{https://bangalore-walkability-index.netlify.app/}}
Grid View displays scores for individual cells, supporting fine-grained analysis
within neighbourhoods, with hover functionality reporting catchment counts per
category at each location. Ward View presents aggregated ward scores for
comparison across BBMP boundaries. Both modes render an interactive heat map
over an OpenStreetMap base layer.

\section{Results and Discussion}
\label{sec:results}

The results below proceed in three steps:
what each personalisation control does to the resulting surface, what happens
when the controls are set to reflect the routines of different kinds of
resident, and how the framework relates to established measures evaluated on the
same input. 

\subsection{Understanding component effects}
\label{sec:results_components}

Each control is examined in isolation, with all others held constant. The figures in this section are rendered rather than captured from the interface. All four figures vary a single control within the demonstration configuration described in Table~\ref{tab:democonfig}, which was designed to make the effect of each control visible.

\begin{table}[htbp]
\caption{The demonstration configuration used for
Figures~\ref{fig:comp_decay}--\ref{fig:comp_tau}. Illustrative only. Reach is the share of
ward cells from which the group is reachable at its own threshold.}
\label{tab:democonfig}
\footnotesize
\begin{tabular*}{\textwidth}{@{\extracolsep{\fill}}lllrr@{}}
\toprule
\textbf{Group} & \textbf{Categories} & \textbf{Treatment} & \textbf{$\tau$} & \textbf{Reach} \\
\midrule
Parks & park & Wanted, standard & 9 & 50.9\,\% \\
Healthcare & hospital, pharmacy & Wanted, standard & 9 & 45.9\,\% \\
Dining & restaurant, fast food, café, & Wanted, standard & 9 & 51.8\,\% \\
 & ice cream, bakery, food court & & & \\
Bars and pubs & bar, pub & Wanted, standard & 12 & 23.3\,\% \\
\bottomrule
\end{tabular*}
\end{table}
\FloatBarrier

\subsubsection{How many nearby}
The decay parameter governs how quickly additional amenities of the same type
stop contributing, and its effect is bounded by how much redundancy is present
to begin with. Replacing the saturating function with simple presence, holding
weights, groups and thresholds fixed, leaves cell-level rankings almost
unchanged across the city ($\rho = 0.997$, $n = 11\,752$) but separates them in
the densest decile ($\rho = 0.952$). Divergence is monotone in redundancy rather
than in threshold: widening the threshold more than doubles the mean count while
moving decay-versus-presence agreement only from 0.9905 to 0.9899. The
explanation is distributional. Among cases where a category is present at all,
38.5\,\% have exactly one instance and 56.8\,\% have at most two, so across most
of the city the decay function has nothing to saturate.

\begin{figure}[!htb]
\centering
\includegraphics[width=\textwidth]{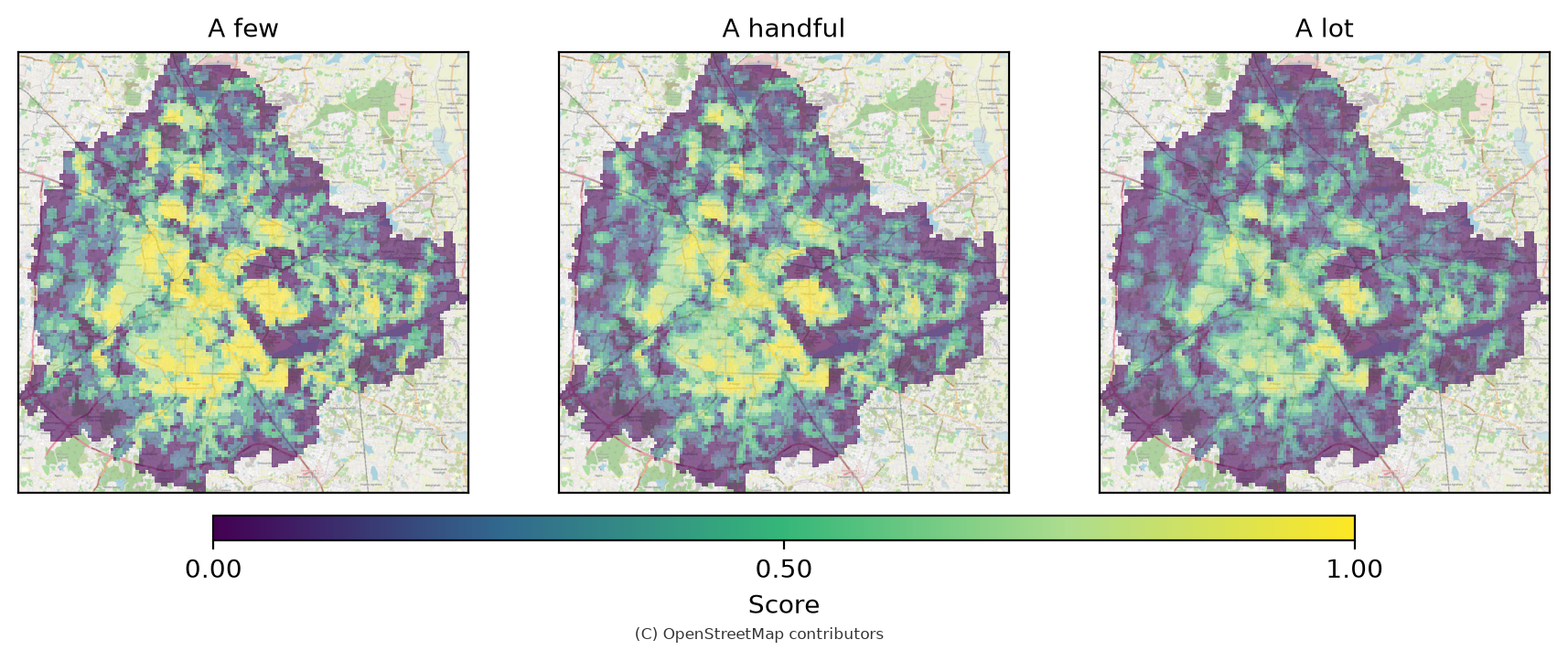}
\caption{The ``how many nearby'' setting varied over a few, a handful and a lot,
applied to every wanted group of the demonstration configuration with weights,
groupings and thresholds held constant.}
\label{fig:comp_decay}
\end{figure}

Figure~\ref{fig:comp_decay} varies this setting across all wanted groups of the
demonstration configuration. \emph{A few} saturates fastest, so a single reachable
amenity already earns most of the available credit and the map is brightest;
\emph{a lot} saturates slowest, so one amenity earns little and the map is
dimmest. The setting changes
how much credit the first instance receives, and therefore how sharply
well-provisioned locations separate from thinly provisioned ones.

\FloatBarrier

\subsubsection{Treatment and weighting}
\begin{figure}[!htb]
\centering
\includegraphics[width=0.82\textwidth]{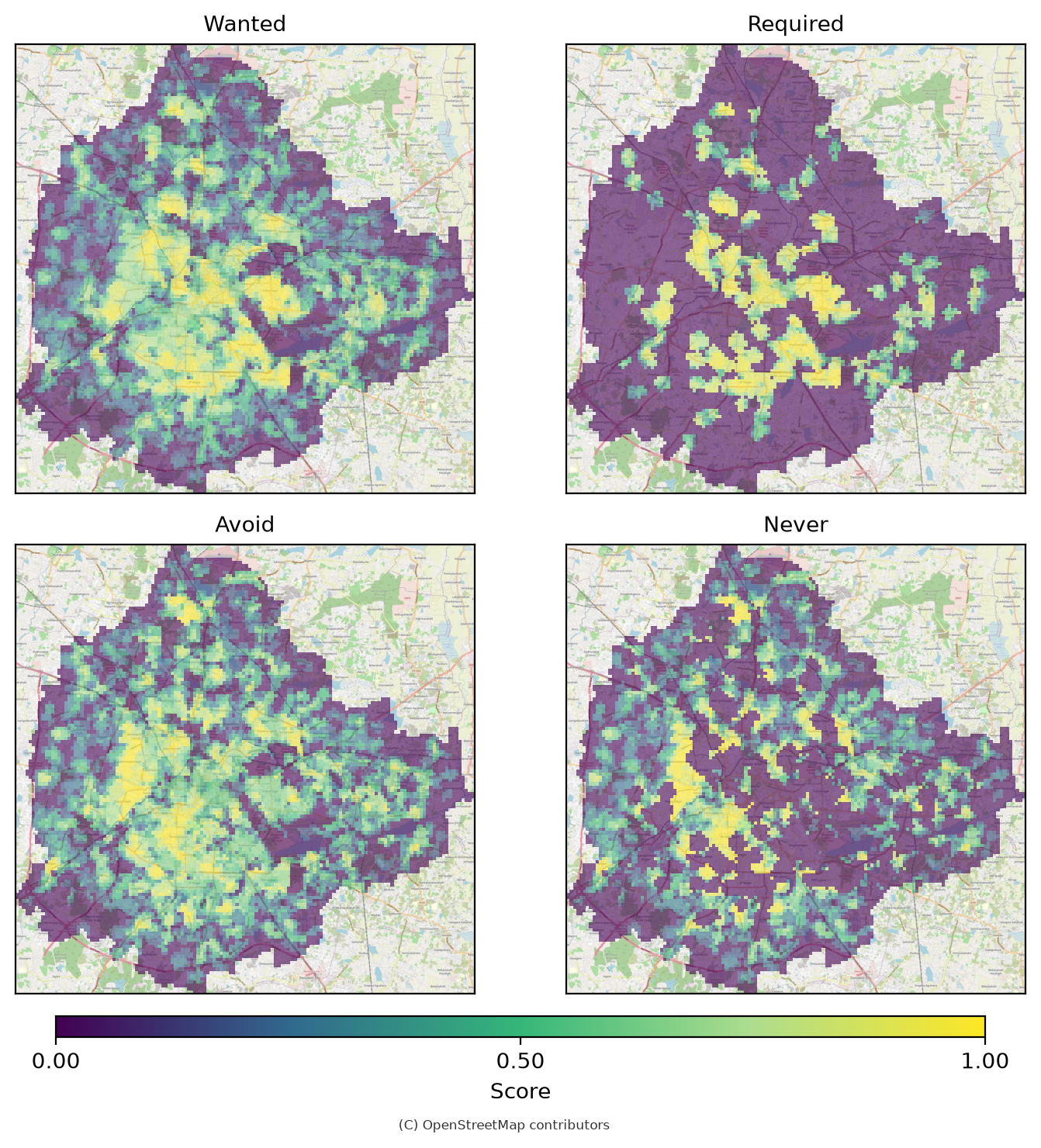}
\caption{The bars-and-pubs group of the demonstration configuration set to each
of the four treatments in turn, everything else held constant.}
\label{fig:comp_treatment}
\end{figure}

Weighting is graded and applies only to Wanted and Avoid, rescaling a category's
contribution without changing what kind of statement the user is making.
Treatment is categorical, fixing which of four statements that is, and
Figure~\ref{fig:comp_treatment} applies all four to the same group so the
differences are visible on one surface.

Under \emph{Wanted} the group contributes like any other, and the map is the
broad provision surface of the base configuration. \emph{Required} makes the
aggregation non-compensatory: every cell without a reachable bar or pub falls
to zero however well it is served otherwise, and what survives is the sparse
set of places that satisfy the constraint, concentrated along the central
corridors.

The lower row inverts both. \emph{Avoid} keeps the whole surface but shades
down the neighbourhoods where the group is present, so the dense centre loses
ground to the periphery while nothing is disqualified outright. \emph{Never}
is its hard form and is the exact complement of Required: the cells Required
retains are precisely those Never removes. Read together, the two rows show
that the distinction is not between more and less weight but between four
different questions.

\FloatBarrier

\subsubsection{Service grouping effects}
\begin{figure}[!htb]
\centering
\includegraphics[width=\textwidth]{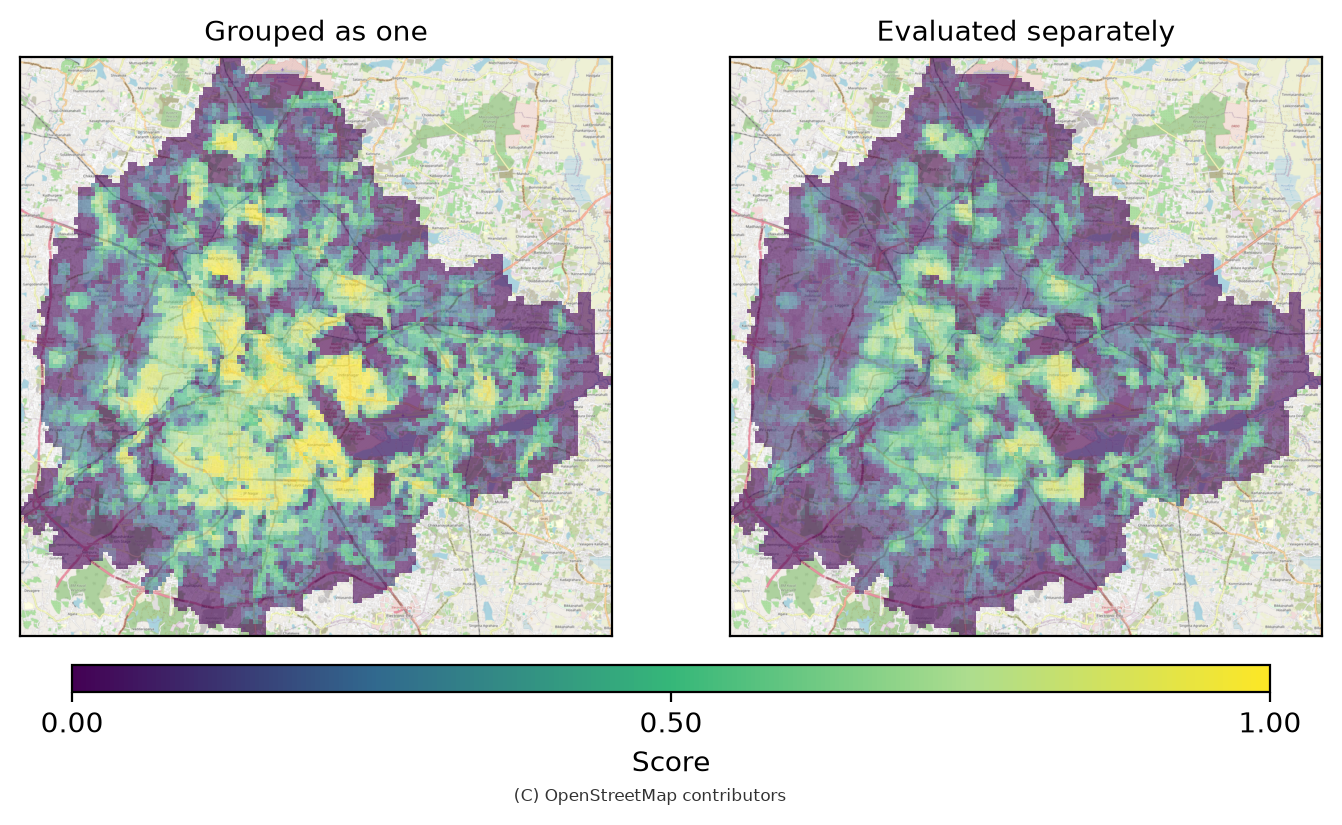}
\caption{The six dining categories of the demonstration configuration evaluated
as one substitutable group and as six independent criteria, everything else held
constant.}
\label{fig:comp_grouping}
\end{figure}

Grouping asserts functional substitutability. A grouped bundle sums its members'
counts before the decay is applied and occupies one criterion's worth of weight;
evaluated separately, each member saturates on its own and each claims its own
share of the denominator, which amounts to asserting that all of them are
independently needed. Figure~\ref{fig:comp_grouping} varies exactly this for the
six dining categories of the demonstration configuration.

The effect is large in level and small in order. Splitting the bundle takes the
denominator from four to nine and drops the mean cell score from 0.331 to 0.212,
lowering 8\,321 of 11\,752 ward cells and raising only 124, because few places
have all six kinds of eating place nearby, so most of the six terms are zero
where the grouped term was satisfied by whichever one was present. Yet cell
rankings barely move ($\rho = 0.989$). The control therefore changes what the
number means far more than which places come out ahead: a user who groups
substitutes is saying that any one of them will do.

\FloatBarrier

\subsubsection{Spatial resolution}
\begin{figure}[!htb]
\centering
\includegraphics[width=\textwidth]{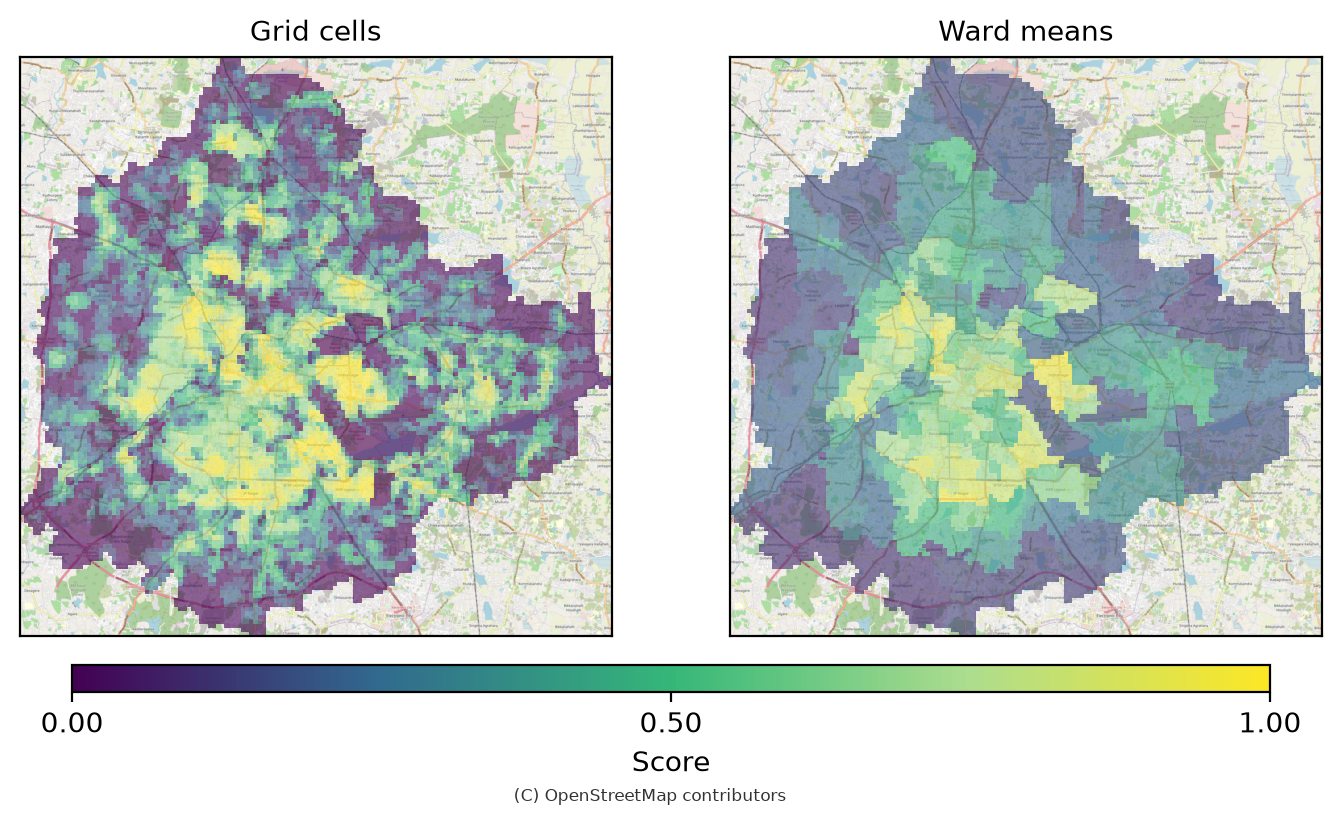}
\caption{One configuration at cell resolution and aggregated to BBMP wards.}
\label{fig:comp_resolution}
\end{figure}

The same configuration answers different questions at cell and at ward level,
and Figure~\ref{fig:comp_resolution} shows both. The cell surface resolves
variation within a neighbourhood, which is what a resident comparing addresses
needs. The ward surface applies Equation~\ref{eq:ward_mean} to those same cells
and aggregates to the unit at which provision is planned and budgets are
allocated.

The cost of aggregation is visible in the pair. Wards whose cells range from
well served to unserved come out uniformly middling, so the sharp local
contrasts of the left panel flatten into a smooth central gradient, and wards
holding large non-residential areas are pulled down by cells nobody lives in.
The ward view is therefore the right unit for comparing places with one another
and the wrong one for choosing an address within a place, which is why every
correlation reported below is computed at ward level while the maps are drawn at
cell level.

\FloatBarrier

\subsubsection{Travel-time threshold effects}
\begin{figure}[!htb]
\centering
\includegraphics[width=0.82\textwidth]{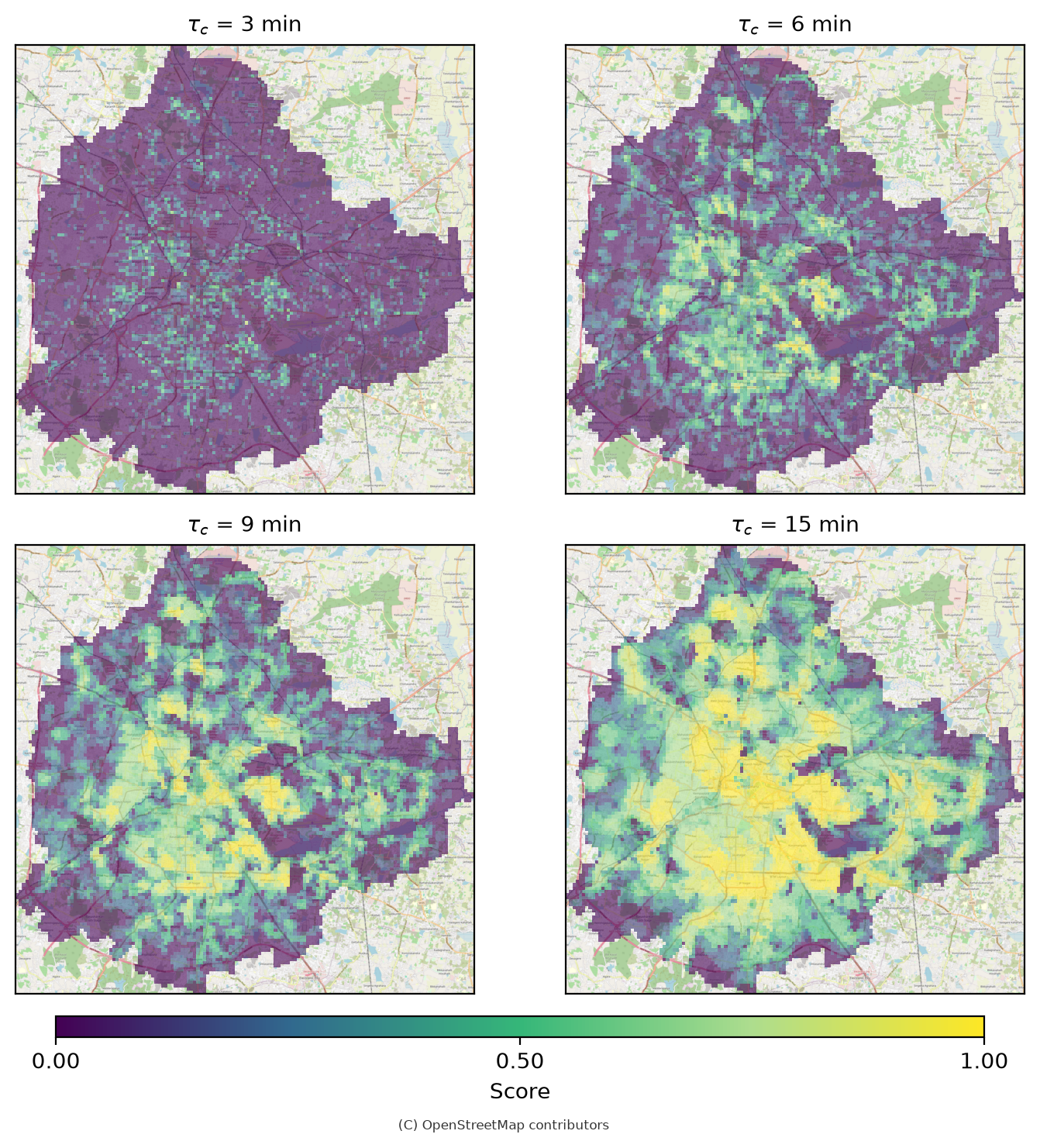}
\caption{A common walking-time threshold of 3, 6, 9 and 15 minutes applied to
every group of the demonstration configuration, everything else held constant.}
\label{fig:comp_tau}
\end{figure}

The threshold determines which amenities enter the calculation at all and has
the largest effect on absolute scores of any control.
Figure~\ref{fig:comp_tau} sets every group of the demonstration configuration to
a common threshold and steps it through four values. At three minutes most of
the area scores near zero; by six the structure of the city appears; by nine it
is legible across the whole area; and by fifteen the centre is close to
uniformly saturated. The progression matters for how the tool is used, because
beyond a certain width the map stops distinguishing between locations: if nearly
everything is reachable from nearly everywhere, category selection carries
little information and every configuration converges on a count of what is
nearby.

\FloatBarrier

\subsection{User profile analysis}
\label{sec:results_profiles}

To show what the controls do together, we built four profiles, one for each of
four kinds of resident, based on which trips dominate their week. We chose trip
purpose because it is the factor best supported by the Bengaluru household
travel survey discussed in Section~\ref{sec:lit_15min}.

\subsubsection{Configuration parameters}
Each profile names its groups and how each is treated; the threshold and decay
of every group are then derived from category-level attributes by a stated rule,
so every value in Table~\ref{tab:profiles} is traceable to that rule and its
source.

For wanted groups the threshold follows the acceptable walking distance
\citet{rahul2021acceptable} report for the trip purpose that motivates the walk,
converted at the walking speed of Section~\ref{sec:method_arch}: 996\,m for work
trips becomes twelve minutes, 666\,m for pick-up and drop-off becomes eight, and
the 360\,m common to recreational, medical and social trips becomes four. The
263\,m reported for personal and household business becomes three, which is the
threshold applied to banking and groceries. Their
analysis finds no significant differences among eat-out, entertainment,
recreation, medical, social and civic purposes, so those are treated as one
bundle rather than split.

Group membership is constrained by coverage rather than chosen freely: a
sparsely mapped member would zero the profile wherever it is missing from the
data rather than absent from the city. Groceries are therefore mapped to
supermarkets alone, despite the argument of Section~\ref{sec:intro_context} that
daily provisioning runs through smaller shops, because general and convenience
stores are under-represented to the degree Section~\ref{sec:results_data}
reports.

Trip purpose cannot set the threshold for something nobody travels to, so
avoided groups take theirs from an exposure radius instead.
\citet{li2022neighborhood} find alcohol outlet density within a 400\,m household
buffer associated with adverse measures in an Indian cohort, and that the
immediate locality matters more than wider availability; 400\,m converts to five
minutes. The international evidence relating outlet density to consumption and
associated harm is reviewed by \citet{campbell2009outlet}.

Required designations appear on schools and parks only. These are the two
categories mapped completely enough to sustain a requirement, for reasons
Section~\ref{sec:results_data} sets out. No profile requires metro access
despite its salience for commuting, because the sparsity of that network is real
rather than an artefact of the data, and requiring it would encode the city's
transit gap as user infeasibility. Avoidance carries no equivalent coverage
floor: an unmapped nuisance produces a penalty of zero, which understates the
problem rather than inventing one, so a sparsely mapped avoided category is
conservative where a sparsely mapped required one would be disqualifying.

\begin{table}[htbp]
\caption{Specification of the four profiles. $\tau$ is in minutes.}
\label{tab:profiles}
\footnotesize
\begin{tabular*}{\textwidth}{@{\extracolsep{\fill}}lllllr@{}}
\toprule
\textbf{Profile} & \textbf{Group} & \textbf{Categories} & \textbf{Treatment} & \textbf{How many} & \textbf{$\tau$} \\
\midrule
Worker & Metro Access & metro station & Wanted, emphasised & a few & 12 \\
 & Dining & restaurant, fast food & Wanted, emphasised & a lot & 4 \\
 & Banking & atm, bank & Wanted, standard & a lot & 3 \\
 & Hospitals & hospital & Wanted, standard & a few & 4 \\
\midrule
School-run & Schools & school & \textbf{Required} & a few & 8 \\
household & Parks & park & Wanted, emphasised & a handful & 4 \\
 & Hospitals & hospital & Wanted, emphasised & a few & 4 \\
 & Groceries & supermarket & Wanted, standard & a handful & 3 \\
 & Banking & bank & Wanted, standard & a handful & 3 \\
 & Alcohol Outlets & bar, pub, alcohol & \textbf{Avoid} & -- & 5 \\
\midrule
Non-worker & Parks & park & \textbf{Required} & a handful & 4 \\
 & Hospitals & hospital & Wanted, emphasised & a few & 4 \\
 & Pharmacies & pharmacy & Wanted, emphasised & a handful & 4 \\
 & Banking & bank, atm & Wanted, standard & a lot & 3 \\
 & Groceries & supermarket & Wanted, standard & a handful & 3 \\
 & Alcohol Outlets & bar, pub, alcohol & \textbf{Avoid}, emphasised & -- & 5 \\
\midrule
Social / & Dining & restaurant, fast food & Wanted, emphasised & a lot & 4 \\
recreational & Nightlife & bar, pub & Wanted, emphasised & a lot & 4 \\
 & Parks & park & Wanted, emphasised & a handful & 4 \\
 & Swimming Pools & swimming pool & Wanted, standard & a few & 4 \\
 & Banking & atm & Wanted, standard & a lot & 3 \\
\bottomrule
\end{tabular*}
\end{table}

\subsubsection{Configuration separates rankings further than formulation does}
\label{sec:results_headline}
\begin{figure}[!htb]
\centering
\includegraphics[width=0.82\textwidth]{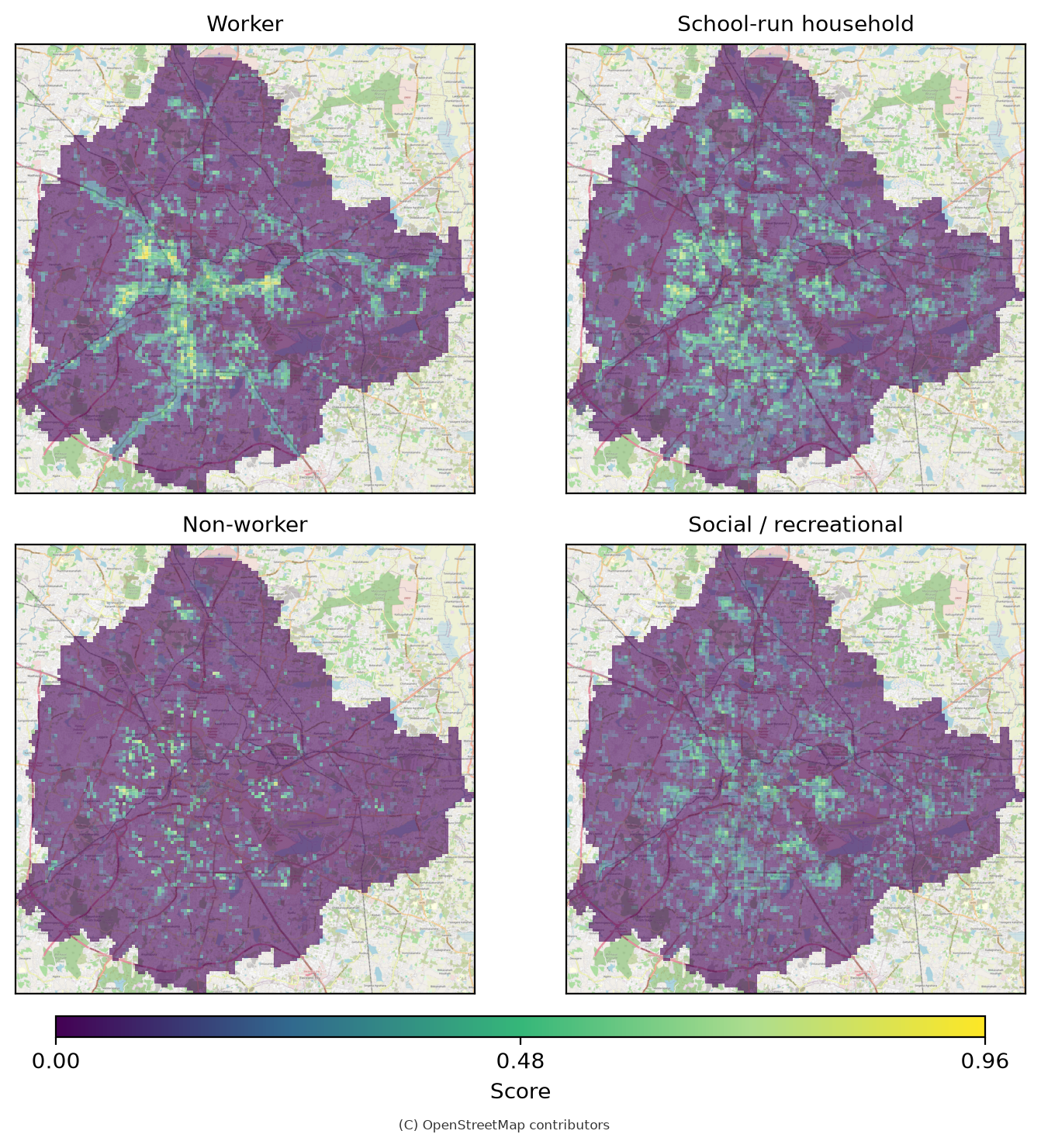}
\caption{Four profiles evaluated across all of BBMP.}
\label{fig:profiles_city}
\end{figure}

Figure~\ref{fig:profiles_city} shows the four profiles across BBMP, and the four
surfaces are not variations on a common pattern. This is the framework's central
empirical result. Measured against a common fixed-preference reference at ward
level, the profiles agree at $\rho = 0.93$, $0.81$, $0.76$ and $0.62$, while the
impedance substitution of Section~\ref{sec:method_formulation} agrees with that
same reference at $\rho = 0.973$. Who is asking separates the rankings
considerably further than the choice of estimator does; the computational
mechanism enables the contribution rather than constituting it.

The profiles also disagree with one another. A binary indicator of whether each
category is present within fifteen minutes reproduces the uniform-weight
configuration at $\rho = 0.997$ but matches the most sharply specified profile
at only $\rho = 0.623$, and the worker and non-worker surfaces agree with each
other at $\rho = 0.585$.

The profiles carrying a Required designation diverge most, which is the expected
consequence of a constraint that alters the feasible set: a ward with excellent
dining, banking and healthcare provision but no reachable school scores zero for
the school-run household, and no amount of other provision compensates. This is
the behaviour a user asking for a non-negotiable need should expect, and it is
not reproducible by weighting alone.

\subsubsection{What avoidance changes}
\label{sec:results_avoid}
Two profiles name alcohol outlets as something to avoid within five minutes,
while a third wants bars and pubs within four, so the same establishments enter
two configurations with opposite sign. The penalty reaches a small part of the
city: 729 of 11\,752 ward cells, 6.2\,\%, have any such outlet within five
minutes, so its aggregate effect is modest and concentrated. It alters 599 cells
for the school-run household and 233 for the non-worker, moving mean scores by
$-4.7\,\%$ and $-10.1\,\%$ and ward rankings by $\rho = 0.996$ and $0.989$
respectively. The larger movement for the non-worker follows from the doubled
weight that profile places on avoidance.

The interesting comparison is not the magnitude but the sign. Domlur is the
highest-scoring ward for the social/recreational profile and simultaneously the
ward whose score the non-worker profile reduces most, by 0.072; Hoysala Nagar
appears near the top of the first list and among the three most penalised in the
second. The two profiles agree at only $\rho = 0.813$ with five of ten top wards
in common, and the amenities responsible largely coincide. A measure that treats
amenity presence as unambiguously good cannot express this, and any single
averaged score for these wards is answering one of the two questions while
appearing to answer both.

\subsection{Comparison with established measures}
\label{sec:results_comparison}

To situate the framework, competing measures were reimplemented on identical
input: the same pedestrian network, amenity set, grid and catchments. Comparing
against scores published by external tools would confound the choice of measure
with differences in underlying data, geography and vintage, so the aggregation
is isolated instead. Baselines are recovered
from the stored count vectors without re-routing, since counts are
non-decreasing in the threshold and differencing them recovers the travel-time
distribution.

\subsubsection{Non-personalised measures}
Table~\ref{tab:baselines} compares four standard measures against this framework
run with uniform weights, so that only the aggregation differs. Run this way the
framework is close to substitutable with much simpler alternatives. A binary
indicator of category presence within fifteen minutes, with no decay and no
parameters, reproduces its ward rankings at $\rho = 0.997$; cumulative
opportunity and a Hansen gravity measure follow at $\rho = 0.985$ and $0.980$.
Rank shift in the table is the median, 90th percentile and maximum number of
places a ward moves out of 198, and Top-20 the number of the twenty
highest-scoring wards shared with this framework.

\begin{table}[htbp]
\caption{Standard measures compared against this framework with uniform weights,
at ward level, on identical routed input.}
\label{tab:baselines}
\footnotesize
\begin{tabular*}{\textwidth}{@{\extracolsep{\fill}}lrrlr@{}}
\toprule
\textbf{Measure} & \textbf{Pearson} & \textbf{Spearman} & \textbf{Rank shift} & \textbf{Top-20} \\
\midrule
Binary presence, 15\,min & 0.997 & 0.997 & 2 / 8 / 13 & 18/20 \\
Cumulative opportunity & 0.931 & 0.985 & 5 / 15 / 36 & 16/20 \\
Gravity (Hansen, $\beta = 0.15$) & 0.918 & 0.980 & 6 / 19 / 44 & 16/20 \\
Walk Score-style reconstruction & 0.950 & 0.956 & 11 / 27 / 66 & 16/20 \\
\bottomrule
\end{tabular*}
\end{table}

Read alongside Section~\ref{sec:results_headline}, this locates the framework's
value precisely. Absent user input the choice of aggregation is nearly
immaterial at ward level; the value lies in the ability to depart from the
uniform default, and that departure moves rankings by an order more than any
choice among these measures.

\subsubsection{A Walk Score-style reconstruction}
\label{sec:results_walkscore}
Walk Score is proprietary and its full weight set unpublished. The comparison
therefore implements the published reconstruction of the Walk Score methodology
given by
\citet{dai2023walkability}, with their published rank weights and
piecewise-linear decay, restricted to the three categories for which weights are
published.

Agreement with our measure is $\rho = 0.946$ across all cells but falls where
provision is dense (Table~\ref{tab:ws_redundancy}). Both measures model
diminishing returns and differ in shape: the published rank weights flatten to
roughly a quarter of the first instance's value by the tenth, whereas the
saturating form decays geometrically toward zero. The orderings therefore
separate only once cells differ in how much redundancy they contain, which is
consistent with the decay analysis in Section~\ref{sec:results_components}.

\begin{table}[htbp]
\caption{Agreement between this framework and the Walk Score-style
reconstruction, conditioned on local amenity density.}
\label{tab:ws_redundancy}
\footnotesize
\begin{tabular*}{\textwidth}{@{\extracolsep{\fill}}lrrr@{}}
\toprule
\textbf{Subset} & \textbf{Cells} & \textbf{Spearman} & \textbf{Mean $k$} \\
\midrule
All cells & 11\,752 & 0.946 & 15.8 \\
Densest 50\,\% & 5\,876 & 0.835 & 29.6 \\
Densest 20\,\% & 2\,350 & 0.461 & 54.0 \\
Densest 10\,\% & 1\,175 & 0.316 & 74.0 \\
\bottomrule
\end{tabular*}
\end{table}

Implementing the reconstruction also required deciding what counts as a grocery
in Bengaluru, and this is where the transferability problem raised in
Section~\ref{sec:intro_context} becomes concrete. Four
alternative mappings were tested. City-wide agreement varied by 0.005 across
them, but dense-area agreement ranged from $\rho = 0.498$ to $\rho = 0.316$,
monotonically with the amount of redundancy each mapping admits. Mapping the
category to supermarkets alone is consistent with the supermarket-oriented
category operationalisation used in the reconstruction and gives
$\rho = 0.498$; mapping it to general and convenience stores reflects how the
city actually provisions itself and gives $\rho = 0.316$. No neutral option
exists: divergence in dense areas is present under every mapping tested, but its
magnitude depends on a judgement about the local service landscape that the
index does not itself supply. \citet{steiniger2019walkmyplace} reach the same
conclusion qualitatively for Chilean cities; the mapping test puts a number on
how much the judgement is worth.

\subsubsection{Measures that accept the same preferences}
\label{sec:results_mcda}
The comparisons above hold preferences fixed, which understates the
alternatives, since weighted linear combination and TOPSIS also accept user
preferences. Each was therefore given the identical categories, weights,
groupings, per-category thresholds \emph{and} negative preferences of each
profile, expressed in its own idiom: an avoided group enters weighted linear
combination as a cost criterion, its normalisation inverted, and enters TOPSIS
as a criterion whose ideal is the column minimum. Both are the textbook
treatments, so no method is handicapped.

The framework differs from a weighted linear combination in two respects: the
saturating normalisation of Equation~\ref{eq:grid_score} in place of a linear
one, and the hard constraints of Equation~\ref{eq:final_score}. Varying them
independently separates their contributions (Table~\ref{tab:mcda}). The
normalisation accounts for 0.020 to 0.059 of ward-rank deviation; the Required
filter accounts for up to 0.246. The two profiles designating no Required group
are rank-identical to this framework with the filter removed ($\rho = 1.000$),
and agree with plain weighted linear combination at $\rho \geq 0.971$.

\begin{table}[htbp]
\caption{Ward-level Spearman agreement with this framework when competing
aggregations are given identical preference data, negative preferences included.
The final column is the share of total deviation attributable to the Required
filter.}
\label{tab:mcda}
\footnotesize
\begin{tabular*}{\textwidth}{@{\extracolsep{\fill}}lrrrrr@{}}
\toprule
\textbf{Profile} & \textbf{WLC} & \textbf{WLC + req.} & \textbf{TOPSIS} & \textbf{Ours, no req.} & \textbf{Req. contributes} \\
\midrule
Worker & 0.971 & 0.971 & 0.990 & 1.000 & 0.000 \\
School-run household & 0.934 & 0.974 & 0.696 & 0.966 & 0.034 \\
Non-worker & 0.749 & 0.941 & 0.541 & 0.754 & 0.246 \\
Social/recreational & 0.980 & 0.980 & 0.982 & 1.000 & 0.000 \\
\bottomrule
\end{tabular*}
\end{table}

Given identical preferences, then, the aggregation behaves like a weighted
linear combination wherever the user designates nothing as essential, and
departs from it exactly where they do. The per-criterion non-compensatory
constraint is what distinguishes the operator, and it is meaningful because the
user rather than the analyst declares it. In the vocabulary of
Section~\ref{sec:lit_mcda}, this is a targeted move toward the conjunctive end
of the ordered weighted averaging spectrum, applied to chosen criteria while the
remainder stay compensatory.

TOPSIS behaves differently. It tracks the
framework closely for the two profiles that name nothing to avoid
($\rho = 0.990$ and $0.982$) and diverges sharply for the two that do
($0.696$ and $0.541$); more than weighted linear combination does on the same
profiles. The cause is that TOPSIS scores an alternative by its relative
distance between the best and worst observed values on each criterion, so a cost
criterion is graded against the worst case anywhere in the city: a ward with no
unwanted outlet scores full marks on that criterion, and one with a single
outlet is graded against the densest concentration in Bengaluru. A subtractive
penalty instead prices each instance on its own terms. Neither is wrong, but
they answer different questions, and the difference only becomes visible once
preferences can point in both directions.

Against the WalkTHERE formulation from which Equation~\ref{eq:grid_score}
extends, evaluated with the same categories and weights, ward-level agreement is
$\rho = 0.973$ and cell-level agreement $\rho = 0.942$. The rectangular
impedance therefore moves rankings little while enabling the precomputation on
which interactivity depends.

\subsection{Data adequacy}
\label{sec:results_data}

Category coverage bounds what such tools can be asked in this setting.
Measuring, for each category, the share of ward cells that cannot reach it even
within fifteen minutes, only two of fifty categories fall below 25\,\%: parks
(23.8\,\%) and schools (24.0\,\%). Forty-three categories exceed 45\,\%. This is
why Required designations appear on only those two categories in
Table~\ref{tab:profiles}.

Some of this is genuine scarcity. Metro stations are unreachable within fifteen
minutes from 86.2\,\% of cells, which reflects the actual extent of the network,
with 74 stations recorded. Other cases reflect the data rather than the city.
General and convenience stores, greengrocers and kiosks together number 806
records across BBMP. Whether these tags are applied exhaustively and mutually
exclusively cannot be established from the extract, so the figure is not a
direct estimate of outlets; but it is implausibly low relative to the scale of
neighbourhood retail in a metropolitan area of this size, where daily
provisioning is dominated by small independent shops. We read the shortfall as
substantial underrepresentation rather than absence.

This bounds the design space directly. An income and care-responsibility axis,
which Section~\ref{sec:intro_context} suggests is a salient distinction in an
Indian city, is not evaluable on this data: a profile requiring daily essentials
returns zero for 96.7\,\% of the city, and relaxing the requirement to a full
fifteen-minute walk still zeroes 53.2\,\%. No threshold repairs a category that
is largely unmapped. Trip purpose is the axis used here, being the one the
present coverage best sustains.

Open amenity data in this setting is markedly more complete for formal,
institutional and commercial amenities than for the informal retail on which
lower-income households rely. A tool built on it is therefore likely to
represent users whose needs are concentrated in well-mapped formal amenities
more faithfully than users who depend on under-mapped informal retail.

\subsection{Individual variation and intended users}
\label{sec:results_audience}

The profiles above are illustrative rather than templates. Individual
preferences within any group vary substantially, and the per-category threshold
is particularly important in accommodating that variation, since two users with
identical weights may reach very different conclusions based on how far each
will walk.

The primary intended user is a resident making a location decision. That is the
use-case the interface is designed around and the one where personalisation is
least problematic: the preferences encoded are the user's own, the consequences
fall on the user, and the alternative otherwise available is a single averaged
score that does not answer the user's particular priorities. The capability is most valuable to
someone moving to an unfamiliar city, where a personalised view reveals service
gaps that a generalised rating conceals.

Two secondary uses follow. Community organisations can characterise service gaps
in terms that matter to the people they represent, without commissioning
technical analysis; here too the person setting the preferences and the people
affected are aligned. Planners and policymakers can examine how provision looks
under different assumptions about whose needs matter, and this is the
consequential case, because it is the one where the person setting the
preferences is not the person who lives with the result.

That second use warrants care. Averaged measures already inform allocation
decisions and encode preferences too, with the disadvantage that those
preferences are implicit and unexaminable. What is specific to a personalised
score is that the person choosing the preferences need not be the person
affected by them. A tool that makes it easy to ask where is good for people like
me also makes it easy to ask which people an area is good for, and the second
question carries different consequences when asked by someone allocating
resources. The same configurability allows a settings choice to be made in
support of a conclusion already reached, which the Required tier makes concrete:
designating one category as non-negotiable can zero large parts of the map. The
concern is sharpened in Bengaluru, where formal and informal provision coexist
within single neighbourhoods, so a tool that scores them differently makes an
existing distinction legible and actionable.

Two safeguards follow, neither costly. Institutional use should record the
configuration alongside the result, so that a score can be interrogated rather
than merely cited. And a personalised score should not serve as the sole
justification for an allocation decision, for the same reason that a single
averaged score should not.

\section{Limitations and Future Work}
\label{sec:limitations}

\subsection{Data quality and coverage}
\label{sec:limitations_data}

Coverage is the framework's principal constraint, and several apparently
separate limitations reduce to it.

Amenity completeness varies by category, as Section~\ref{sec:results_data}
quantifies. Informal retail appears at a small fraction of its likely density
while formal and institutional amenities are represented reasonably, so
configurations depending on informal retail cannot currently be evaluated
meaningfully. This is the sense in which the framework is not preference-neutral
in practice: it is most expressive for users whose needs are met by formal
provision, and least expressive for users who depend on the categories the data
under-represents, for whom walking access is also least substitutable by other
modes.

All instances of a category are treated as equivalent, distinguished only by
count and proximity, because the data carries no capacity or quality attribute.
A single-room clinic and a tertiary hospital contribute identically, as do a
kiosk and a supermarket. This is most consequential where a Required designation
is satisfied by any single facility regardless of whether its capacity matches
the need, and it is also why two-step floating catchment area methods
\citep{luo2003measures} were not adopted: without a capacity attribute, every
provider-to-population ratio would reduce to a facility count.

Walking speed is treated as a single constant, though pedestrian speed varies
with footpath provision, crossing density, gradient, crowding and the walker's
age and mobility. The attributes required to vary it spatially are not
consistently mapped, so a single well-supported constant is preferable to a
spatially varying estimate built on incomplete data.

Relatedly, the framework measures proximity only. The pedestrian network encodes
connectivity, not suitability: infrastructure quality, street lighting, traffic
safety, shade and crossing provision are excluded, which matters where roughly
thirty per cent of roads carry footpaths \citep{rahul2021acceptable}. Two
locations with identical scores may differ considerably in whether the walk is
pleasant or safe.

Finally, ward scores are unweighted spatial means, so wards containing large
non-residential areas show deflated scores. Population weighting requires
sub-ward counts, and the most recent Indian census was taken in 2011;
Bengaluru's population has changed substantially since, so weighting by those
counts would introduce an error of its own rather than remove one.

Integrating a commercial place database would address the greater part of this:
better retail coverage would make the income-differentiated design
Section~\ref{sec:results_data} reports as unevaluable viable, capacity and rating
attributes would permit quality-weighted scoring and open the two-step floating
catchment area family, and richer pedestrian attribution would support spatially
varying speeds. The trade is reproducibility against coverage, and a practical
route is to keep OpenStreetMap as the base while offering commercial sources as
an optional overlay.

\subsection{Parameter defaults and calibration}
\label{sec:limitations_params}

Every control defaults to a common setting (Wanted, Standard weight, a handful
nearby, a nine-minute threshold) and none of those defaults is empirically
derived. The thresholds used in Section~\ref{sec:results_profiles} are informed
by published acceptable distances \citep{rahul2021acceptable}, but those are
population-level summaries from a single household survey conducted in 2012--13,
predating most of Bengaluru's metro network. The decay settings are selected
rather than calibrated, and the rate at which marginal utility declines almost
certainly differs by category; the penalty rate applied to avoided categories is
fixed at the middle setting for the same want of evidence. The two-level
weighting is likewise a usability compromise rather than a measured preference
structure.

The framework treats these as quantities the user supplies, which is appropriate
for its purpose but leaves an open question that a different study could answer:
what settings should a user who has no particular view adopt? A stated- or
revealed-preference study could estimate category-specific decay rates,
purpose-specific thresholds and relative weights together, yielding defaults
with an empirical basis and, more usefully, an indication of how much any
individual is likely to depart from them. Collecting behavioural data for that
purpose is the most valuable single extension of this work.

\subsection{Scope of validation and evaluation}
\label{sec:limitations_validation}

The contribution is a tool and architecture rather than a validated empirical
measurement of accessibility in Bengaluru. Construct validation, establishing
that reported scores correspond to accessibility as residents experience it,
would require an external reference that the uneven coverage described above
makes unavailable, since any disagreement would conflate genuine accessibility
differences with differences in mapping. What is reported instead is
method-level validation: agreement with independently routed catchments,
sensitivity to every discretisation parameter, and comparison against
established measures on identical input.

No usability study has been conducted, so
whether users understand the controls, and whether their configurations reflect
their actual preferences, are open questions.

Coverage is estimated by sampling rather than exact area computation, so the
estimand is the fraction of a cell whose nearest network node is reachable
rather than true point reachability, which is immaterial for scores but a
residual source of disagreement in individual counts near the coverage
threshold.

\subsection{Non-technical limitations}
\label{sec:limitations_social}

Section~\ref{sec:results_audience} sets out who the framework is for and why
institutional use is the consequential case. Two consequences of that use lie
beyond what it can represent. If provision were repeatedly optimised against the
aggregate of personalised configurations, an area would specialise toward the
profiles already over-represented among its residents, reinforcing residential
sorting; the framework evaluates one configuration at one instant and has no
dynamics, so it can neither represent that process nor detect it. And a score
functions as a signal as much as a measurement: a high value under an affluent
configuration can accelerate displacement, a low value under a widely used one
can rationalise disinvestment. The measure reports what is reachable, not who
should benefit, and the safeguards proposed there are procedural rather than
substantive. They are also untested: whether recording configurations would make
institutional scores interrogable, and how planners would actually configure the
tool, are empirical questions about institutions rather than about software.

A further limit is representational. The framework expresses only preferences
that take the form of amenity categories, so needs met relationally or informally
(childcare through kin networks, credit through local relationships, facilities
specific to a community) are invisible to it however complete the amenity data
becomes.

\section{Conclusions}
\label{sec:conclusions}

Accessibility research has long held that accessibility differs between
individuals sharing a location, and supplies measures that represent those
differences; those measures are run offline by analysts. Public walkability
tools are immediate and place-based, returning one value to everyone at a
location. This work implements and deploys a framework in which the user sets
the preferences and the map answers at once, across all 198 wards of Bengaluru.
Three findings follow, with two qualifications.

Preference specification separates rankings further than the choice of
aggregation does. Against a common fixed-preference reference the four profiles
agree at between $\rho = 0.93$ and $\rho = 0.62$, where the impedance
substitution that distinguishes our formulation agrees at $\rho = 0.973$, and
where uniform-weight runs of four standard measures agree at $\rho \geq 0.956$.

Hard constraints produce the largest divergence from compensatory aggregation.
Given identical preferences the framework tracks a weighted linear combination
closely unless a category is marked Required or Never, which contributes up to
$0.246$ of ward-rank deviation against $0.020$--$0.059$ for the normalisation.
The reason is that these designations change which locations are feasible rather
than how feasible locations rank, which no choice of weight reproduces.

The computational simplification preserves rankings while enabling interaction.
A rectangular impedance collapses the inner summation to a count, so routing and
intersection happen once offline and a reconfiguration costs one lookup and a
little arithmetic per category; against the continuous-impedance formulation it
extends, ward-level agreement is $\rho = 0.973$. This substitution is what
converts a batch measure into an interactive one.

Two qualifications bound these findings. Results are robust to discretisation:
grid resolution, coverage threshold and sampling density each shift absolute
scores by a few per cent while leaving ward rankings essentially unchanged. But
open data coverage bounds who the measure can represent. Amenity data in
Bengaluru is markedly more complete for formal provision than for informal
retail, to the point where a profile differentiated by income and care
responsibility returns zero for 96.7\,\% of the city.

That last point sets the terms. Work of this kind is usually placed under
Sustainable Development Goal 11 of the United Nations 2030 Agenda
\citep{un2030agenda}, whose targets on access to public transport (11.2) and to
green and public space (11.7) are themselves cumulative proximity measures,
evaluated with thresholds and category definitions set centrally rather than by
the residents being counted. A configurable measure does serve inclusivity by
making differences between residents explicit rather than averaging them away. But it represents only the needs it has data for, and on
present coverage it describes needs met by formal provision more faithfully than
those met informally. The gap is a property of the data rather than the method
and is tractable; until it closes it is better stated than concealed behind a
single figure.

\section*{Declarations}

\bmhead{Ethical Approval}
Not applicable. This study did not involve human participants or animals.

\bmhead{Consent to Participate}
Not applicable.

\bmhead{Consent to Publish}
Not applicable.

\bmhead{Funding}
This research received no specific grant from any funding agency in the public,
commercial, or not-for-profit sectors.

\bmhead{Competing Interests}
The authors declare that they have no competing interests.

\bmhead{Data Availability}
All amenity data used in this study are derived from OpenStreetMap and are
publicly available. The pedestrian network was built from an OpenStreetMap
extract \citep{osm_france_india}, and ward boundaries were obtained from the
Urban Data Portal \citep{urban_data_portal}. The interactive application
implementing the framework, including the derived accessibility scores, is
publicly accessible at
\url{https://bangalore-walkability-index.netlify.app/}.

\bmhead{Code Availability}
The scoring engine, the profile construction, the comparisons reported here and
the precomputed data they read are available at
\url{https://github.com/GPrabhanjana/personalised-urban-accessibility}.

\bmhead{Author Contributions}
Prabhanjana Ghuriki conceptualised the study, developed the methodology,
implemented the computational framework and web application, conducted the
spatial analysis, collected data, and drafted the manuscript. S Chanti
contributed to the literature review and manuscript revision. Both authors read
and approved the final manuscript.

\bibliography{references}

@article{xu2020deconstructing,
author = {Yanyan Xu  and Luis E. Olmos  and Sofiane Abbar  and Marta C. González },
title = {Deconstructing laws of accessibility and facility distribution in cities},
journal = {Science Advances},
volume = {6},
number = {37},
pages = {eabb4112},
year = {2020},
doi = {10.1126/sciadv.abb4112},
URL = {https://www.science.org/doi/abs/10.1126/sciadv.abb4112},
eprint = {https://www.science.org/doi/pdf/10.1126/sciadv.abb4112}
}

@article{sudhira2007urban,
author = {H S, Sudhira and Ramachandra, T V},
year = {2007},
month = {01},
pages = {},
title = {Characterising urban sprawl from remote sensing data and using landscape metrics},
journal = {Proceedings of 10th International Conference on Computers in Urban Planning and Urban Management, CUPUM 2007}
}

@article{shaw2012urban,
author = {Annapurna Shaw},
title = {Metropolitan City Growth and Management in Post-liberalized India},
journal = {Eurasian Geography and Economics},
volume = {53},
number = {1},
pages = {44--62},
year = {2012},
publisher = {Routledge},
doi = {10.2747/1539-7216.53.1.44},
URL = {https://doi.org/10.2747/1539-7216.53.1.44},
eprint = {https://doi.org/10.2747/1539-7216.53.1.44}
}

@book{jacobs1961death,
  title = {The Death and Life of Great American Cities},
  year = {1961},
  publisher = {Vintage Books},
  author = {Jacobs, Jane}
}

@article{geurs2004,
title = {Accessibility evaluation of land-use and transport strategies: review and research directions},
journal = {Journal of Transport Geography},
volume = {12},
number = {2},
pages = {127-140},
year = {2004},
issn = {0966-6923},
doi = {https://doi.org/10.1016/j.jtrangeo.2003.10.005},
url = {https://www.sciencedirect.com/science/article/pii/S0966692303000607},
author = {Karst T. Geurs and Bert {van Wee}}
}

@article{ingram1971,
author = {D.R. Ingram},
title = {The concept of accessibility: A search for an operational form},
journal = {Regional Studies},
volume = {5},
number = {2},
pages = {101--107},
year = {1971},
publisher = {RSA Website},
doi = {10.1080/09595237100185131},
}

@article{hansen1959,
author = {Walter G. Hansen},
title = {How Accessibility Shapes Land Use},
journal = {Journal of the American Institute of Planners},
volume = {25},
number = {2},
pages = {73--76},
year = {1959},
publisher = {Routledge},
doi = {10.1080/01944365908978307},
URL = {https://doi.org/10.1080/01944365908978307},
eprint = {https://doi.org/10.1080/01944365908978307}
}

@article{carr2010walk,
author = {Carr, Lucas and Dunsiger, Shira and Marcus, Bess},
year = {2010},
month = {11},
pages = {460-3},
title = {Walk Score (TM) As a Global Estimate of Neighborhood Walkability},
volume = {39},
journal = {American journal of preventive medicine},
doi = {10.1016/j.amepre.2010.07.007}
}

@unknown{nattererparis,
author = {Natterer, Elena and Loder, Allister and Bogenberger, Klaus},
year = {2023},
month = {08},
pages = {},
title = {Traffic Reduction and Decarbonization through Network Changes - Empirical Evidence from Paris}
}

@article{BUTTNER2024100095,
title = {Just around the corner: Accessibility by proximity in the 15-minute city},
journal = {Journal of Urban Mobility},
volume = {6},
pages = {100095},
year = {2024},
issn = {2667-0917},
doi = {https://doi.org/10.1016/j.urbmob.2024.100095},
url = {https://www.sciencedirect.com/science/article/pii/S2667091724000256},
author = {Benjamin Büttner and Cecília Silva and Louis Merlin and Karst Geurs}
}

@article{NIEUWENHUIJSENbarcelona,
title = {The Superblock model: A review of an innovative urban model for sustainability, liveability, health and well-being},
journal = {Environmental Research},
volume = {251},
pages = {118550},
year = {2024},
issn = {0013-9351},
doi = {https://doi.org/10.1016/j.envres.2024.118550},
url = {https://www.sciencedirect.com/science/article/pii/S0013935124004547},
author = {Mark Nieuwenhuijsen and Audrey {de Nazelle} and Marta Cirach Pradas and Carolyn Daher and Angel M. Dzhambov and Cynthia Echave and Stefan Gössling and Tamara Iungman and Haneen Khreis and Nicolina Kirby and Sasha Khomenko and Ulrich Leth and Florian Lorenz and Vlatka Matkovic and Johannes Müller and Laia Palència and Evelise {Pereira Barboza} and Katherine Pérez and Lambed Tatah and Jernej Tiran and Cathryn Tonne and Natalie Mueller}
}

@book{buttner2024mapping,
author = {Büttner, Benjamin and Seisenberger, Sebastian and Mccormick, Bartosz and Silva, Cecília and Teixeira, João and Papa, Enrica},
year = {2024},
month = {09},
pages = {},
title = {Mapping of 15-minute City Practices: Overview on strategies, policies and implementation in Europe and beyond},
doi = {10.13140/RG.2.2.12661.79842}
}

@Article{smartcities,
AUTHOR = {Moreno, Carlos and Allam, Zaheer and Chabaud, Didier and Gall, Catherine and Pratlong, Florent},
TITLE = {Introducing the “15-Minute City”: Sustainability, Resilience and Place Identity in Future Post-Pandemic Cities},
JOURNAL = {Smart Cities},
VOLUME = {4},
YEAR = {2021},
NUMBER = {1},
PAGES = {93--111},
URL = {https://www.mdpi.com/2624-6511/4/1/6},
ISSN = {2624-6511},
DOI = {10.3390/smartcities4010006}
}

@misc{osm_france_india,
  author       = {{OpenStreetMap France}},
  title        = {Asia / India Extracts},
  howpublished = {\url{https://download.openstreetmap.fr/extracts/asia/india/}},
  note         = {Accessed: 2025-07-15},
  year         = {2025}
}

@Article{ijgi13120450,
AUTHOR = {Mazzulla, Gabriella and Pirrone, Carlo Giuseppe},
TITLE = {Accessibility Measures: From a Literature Review to a Classification Framework},
JOURNAL = {ISPRS International Journal of Geo-Information},
VOLUME = {13},
YEAR = {2024},
NUMBER = {12},
ARTICLE-NUMBER = {450},
URL = {https://www.mdpi.com/2220-9964/13/12/450},
ISSN = {2220-9964},
DOI = {10.3390/ijgi13120450}
}

@article{CURL20113,
title = {Does Accessibility Planning address what matters? A review of current practice and practitioner perspectives},
journal = {Research in Transportation Business \& Management},
volume = {2},
pages = {3-11},
year = {2011},
note = {Accessibility in passenger transport: policy and management},
issn = {2210-5395},
doi = {https://doi.org/10.1016/j.rtbm.2011.07.001},
url = {https://www.sciencedirect.com/science/article/pii/S2210539511000204},
author = {Angela Curl and John D. Nelson and Jillian Anable}
}

@misc{albashir_2024_14231533,
  author       = {Albashir, Abubakr and
                  Messa, Federico and
                  Presicce, Dante and
                  Pedrazzoli, Alessandro and
                  Gorrini, Andrea},
  title        = {15min City Score Toolkit – Urban Walkability
                   Analytics
                  },
  month        = nov,
  year         = 2024,
  publisher    = {Zenodo},
  doi          = {10.5281/zenodo.14231533},
  url          = {https://doi.org/10.5281/zenodo.14231533},
}

@article{Brownetal,
author = {Jennifer Ann Brown and Kimberley D. Curtin and Mathew Thomson and Janice Y. Kung and Candace I. J. Nykiforuk},
title ={Contributions and Limitations Walk Score® in the Context of Walkability: A Scoping Review},
journal = {Environment and Behavior},
volume = {55},
number = {6-7},
pages = {468-519},
year = {2023},
doi = {10.1177/00139165231201611},
URL = {https://doi.org/10.1177/00139165231201611},
eprint = {https://doi.org/10.1177/00139165231201611}
}

@article{neutens2010,
  author       = {Neutens, Tijs and Schwanen, Tim and Witlox, Frank and De Maeyer, Philippe},
  issn         = {0308-518X},
  journal      = {Environment and Planning A},
  number       = {7},
  pages        = {1613--1635},
  title        = {Equity of urban service delivery: a comparison of different accessibility measures},
  url          = {http://doi.org/10.1068/a4230},
  volume       = {42},
  year         = {2010},
}

@article{Fadda2024,
title = {How to locate services optimizing redundancy: A comparative analysis of K-Covering Facility Location models},
journal = {Socio-Economic Planning Sciences},
volume = {94},
pages = {101938},
year = {2024},
issn = {0038-0121},
doi = {https://doi.org/10.1016/j.seps.2024.101938},
url = {https://www.sciencedirect.com/science/article/pii/S003801212400137X},
author = {Edoardo Fadda and Daniele Manerba and Roberto Tadei}
}

@article{Levinson_Wu_2020, title={Towards a general theory of access}, volume={13}, url={https://www.jtlu.org/index.php/jtlu/article/view/1660}, DOI={10.5198/jtlu.2020.1660}, abstractNote={&amp;lt;p&amp;gt;This paper integrates and extends many of the concepts of accessibility deriving from Hansen’s (1959) seminal paper, and develops a theory of access that generalizes from the particular measures of access that have become increasingly common. Access is now measured for a particular place by a particular mode for a particular purpose at a particular time in a particular year. General access is derived as a theoretical ideal that would be measured for all places, all modes, all purposes, at all times, over the lifecycle of a project. It is posited that more general access measures better explain spatial location phenomena.&amp;lt;/p&amp;gt;}, number={1}, journal={Journal of Transport and Land Use}, author={Levinson, David M and Wu, Hao}, year={2020}, month={Jun.}, pages={129–158} }

@article{roper2023incorporating,
author = {Roper, Josephine and Ng, Matthew and Pettit, Christopher},
year = {2023},
month = {10},
pages = {361-387},
title = {Incorporating diminishing returns to opportunities in access: Development of an open-source walkability index based on multi-activity accessibility},
volume = {16},
journal = {Journal of Transport and Land Use},
doi = {10.5198/jtlu.2023.2308}
}

@article{JIANG2024103197,
title = {Mapping liveability: The “15-min city” concept for car-dependent districts in Auckland, New Zealand},
journal = {Applied Geography},
volume = {163},
pages = {103197},
year = {2024},
issn = {0143-6228},
doi = {https://doi.org/10.1016/j.apgeog.2024.103197},
url = {https://www.sciencedirect.com/science/article/pii/S014362282400002X},
author = {Jing Jiang and Weijie Qiao and I-Ting Chuang and Yan Li and Tianyi Wang and Lee Beattie}
}

@misc{valhalla_routing,
  title        = {Valhalla},
  author       = {{Valhalla Contributors}},
  howpublished = {\url{https://github.com/valhalla/valhalla}},
  version      = {v3.5.1},
  year         = {2024},
  note         = {Accessed: July 2025}
}

@misc{urban_data_portal,
  author = {{Urban Data Portal}},
  title = {{BBMP} Ward Map - 2015},
  year = {2015},
  howpublished = {Dataset},
  publisher = {OpenCity},
  url = {https://data.opencity.in/dataset/bbmp-ward-information/resource/bbmp-ward-map---2015},
  note = {Accessed: 2025-09-29}
}

@article{rahul2021acceptable,
  author  = {Rahul, T. M. and Manoj, M. and Tahlyan, Divyakant and Verma, Ashish},
  title   = {The Influence of Activities and Socio-Economic/Demographic Factors on the Acceptable Distance in an {I}ndian Scenario},
  journal = {Transportation in Developing Economies},
  volume  = {7},
  number  = {1},
  pages   = {3},
  year    = {2021},
  doi     = {10.1007/s40890-020-00110-3}
}

@article{banerjee2018pedestrian,
  author  = {Banerjee, Arunabha and Maurya, Akhilesh Kumar and L{\"a}mmel, Gregor},
  title   = {A Review of Pedestrian Flow Characteristics and Level of Service over Different Pedestrian Facilities},
  journal = {Collective Dynamics},
  volume  = {3},
  pages   = {1--52},
  year    = {2018},
  doi     = {10.17815/CD.2018.17}
}

@inproceedings{dai2023walkability,
  author    = {Dai, Weimin and Yin, Cong and Khalil, Elias B. and Sanner, Scott},
  title     = {Walkability Optimization: Formulations, Algorithms, and a Case Study of Toronto},
  booktitle = {arXiv preprint},
  year      = {2022},
  eprint    = {2212.05192},
  archivePrefix = {arXiv},
  url       = {https://arxiv.org/abs/2212.05192},
  note      = {Formalisation of the Walk Score methodology used in this paper}
}

@article{malczewski2006gis,
  author  = {Malczewski, Jacek},
  title   = {{GIS}-based multicriteria decision analysis: a survey of the literature},
  journal = {International Journal of Geographical Information Science},
  volume  = {20},
  number  = {7},
  pages   = {703--726},
  year    = {2006},
  doi     = {10.1080/13658810600661508}
}

@article{yager1988owa,
  author  = {Yager, Ronald R.},
  title   = {On ordered weighted averaging aggregation operators in multicriteria decisionmaking},
  journal = {IEEE Transactions on Systems, Man, and Cybernetics},
  volume  = {18},
  number  = {1},
  pages   = {183--190},
  year    = {1988},
  doi     = {10.1109/21.87068}
}

@misc{henry2024close,
  author       = {Henry, Nathaniel},
  title        = {Close: An Interactive, Multi-Modal Travel Time Map},
  year         = {2024},
  howpublished = {Henry Spatial Analysis},
  url          = {https://close.city/},
  note         = {Accessed: 2026-08}
}

@misc{epa2021walkability,
  author       = {{U.S. Environmental Protection Agency}},
  title        = {National Walkability Index: User Guide and Methodology},
  year         = {2021},
  howpublished = {Office of Community Revitalization},
  url          = {https://www.epa.gov/smartgrowth/smart-location-mapping},
  note         = {Accessed: 2026-08}
}

@article{luo2003measures,
  author  = {Luo, Wei and Wang, Fahui},
  title   = {Measures of spatial accessibility to health care in a {GIS}
             environment: synthesis and a case study in the {Chicago} region},
  journal = {Environment and Planning B: Planning and Design},
  volume  = {30},
  number  = {6},
  pages   = {865--884},
  year    = {2003},
  doi     = {10.1068/b29120}
}

@article{li2022neighborhood,
  author  = {Li, Yingjun and Mallinson, Poppy Alice Carson and Aggarwal, Aastha
             and Kulkarni, Bharati and Kinra, Sanjay},
  title   = {Association of neighborhood alcohol environment with alcohol intake
             and cardiovascular risk factors in {India}: cross-sectional evidence
             from {APCAPS}},
  journal = {Frontiers in Cardiovascular Medicine},
  volume  = {9},
  pages   = {844086},
  year    = {2022},
  doi     = {10.3389/fcvm.2022.844086}
}

@article{campbell2009outlet,
  author  = {Campbell, Carla Alexia and Hahn, Robert A. and Elder, Randy and
             Brewer, Robert and Chattopadhyay, Sajal and Fielding, Jonathan and
             Naimi, Timothy S. and Toomey, Traci and Lawrence, Briana and
             Middleton, Jennifer Cook},
  title   = {The effectiveness of limiting alcohol outlet density as a means of
             reducing excessive alcohol consumption and alcohol-related harms},
  journal = {American Journal of Preventive Medicine},
  volume  = {37},
  number  = {6},
  pages   = {556--569},
  year    = {2009},
  doi     = {10.1016/j.amepre.2009.09.028}
}

@article{kwan1998space,
  author  = {Kwan, Mei-Po},
  title   = {Space-time and integral measures of individual accessibility:
             a comparative analysis using a point-based framework},
  journal = {Geographical Analysis},
  volume  = {30},
  number  = {3},
  pages   = {191--216},
  year    = {1998},
  doi     = {10.1111/j.1538-4632.1998.tb00396.x}
}

@article{pot2021perceived,
  author  = {Pot, Felix J. and van Wee, Bert and Tillema, Taede},
  title   = {Perceived accessibility: what it is and why it differs from
             calculated accessibility measures based on spatial data},
  journal = {Journal of Transport Geography},
  volume  = {94},
  pages   = {103090},
  year    = {2021},
  doi     = {10.1016/j.jtrangeo.2021.103090}
}

@article{klumpenhouwer2021bundling,
  author  = {Klumpenhouwer, Willem and Huang, Wanyang},
  title   = {A flexible framework for measuring accessibility with
             destination bundling},
  journal = {Journal of Transport Geography},
  volume  = {91},
  pages   = {102949},
  year    = {2021},
  doi     = {10.1016/j.jtrangeo.2021.102949}
}

@inproceedings{steiniger2019walkmyplace,
  author    = {Steiniger, Stefan and Rojas, Carolina and Vecchio, Giovanni},
  title     = {Evaluating walkability with {WalkMyPlace}: five reasons why the
               utility of {Walkscore.com} may be limited in {South America}},
  booktitle = {Accepted Short Papers and Posters, 22nd AGILE Conference on
               Geographic Information Science},
  address   = {Limassol, Cyprus},
  pages     = {17--20},
  year      = {2019}
}

@misc{un2030agenda,
  author       = {{United Nations General Assembly}},
  title        = {Transforming our World: The 2030 Agenda for Sustainable Development},
  year         = {2015},
  note         = {Resolution A/RES/70/1},
  url          = {https://sdgs.un.org/2030agenda}
}

\end{document}